\PassOptionsToPackage{table}{xcolor}
\documentclass[unnumsec,webpdf,contemporary,large]{oup-authoring-template}

\usepackage{microtype}
\usepackage{graphicx}
\usepackage{flushend} 

\usepackage{booktabs}
\usepackage{adjustbox}
\usepackage{array}
\usepackage[table]{xcolor}
\usepackage{colortbl}
\usepackage{multirow}
\usepackage{placeins} 
\usepackage{float} 

\usepackage{amsmath}
\usepackage{amssymb}
\usepackage{mathtools}
\usepackage{bm}

\usepackage{pifont}
\usepackage{xparse}
\newcommand{\cmark}{\ding{51}}
\newcommand{\xmark}{\ding{55}}

\AtBeginDocument{%
  \hypersetup{colorlinks=true,citecolor=blue,linkcolor=black,urlcolor=blue}%
  \let\origcitealp\citealp
  \RenewDocumentCommand{\citep}{O{} O{} m}{(\textcolor{blue}{\origcitealp[#1][#2]{#3}})}%
}

\definecolor{HeadPurple}{RGB}{235,228,255}

\begin{document}

\journaltitle{}
\DOI{}
\copyrightyear{}
\pubyear{}
\vol{}
\issue{}
\access{}
\appnotes{}

\firstpage{1}

\title[BRIDGE for gene regulatory network inference]{\begingroup\fontsize{18bp}{22bp}\selectfont BRIDGE: Biological Evidence Refinement and Heterogeneous Dynamic Gating for Gene Regulatory Networks\endgroup}

\author[]{%
\parbox{\textwidth}{\centering
Ziyang Dong\textsuperscript{1,\textdagger},
Shanwen Tan\textsuperscript{1,\textdagger},
Hengchuang Yin\textsuperscript{2},
Wei Liu\textsuperscript{3},
Yifan Wang\textsuperscript{4},\\
Siyu Yi\textsuperscript{3,*},
Jiancheng Lv\textsuperscript{1}
and Wei Ju\textsuperscript{5,*}
}}
\address[1]{\orgdiv{College of Computer Science}, \orgname{Sichuan University}, \orgaddress{\state{Chengdu}, \postcode{610065}, \country{China}}}
\address[2]{\orgname{Xinjiang Technical Institute of Physics and Chemistry, Chinese Academy of Sciences}, \orgaddress{\state{Urumqi}, \postcode{830011}, \country{China}}}
\address[3]{\orgdiv{School of Mathematics}, \orgname{Sichuan University}, \orgaddress{\state{Chengdu}, \postcode{610065}, \country{China}}}
\address[4]{\orgname{School of Artificial Intelligence and Data Science, University of International Business and Economics}, \orgaddress{\state{Beijing}, \postcode{100029}, \country{China}}}
\address[5]{\orgdiv{School of Artificial Intelligence}, \orgname{Sichuan University}, \orgaddress{\state{Chengdu}, \postcode{610065}, \country{China}}}

\corresp[*]{Corresponding authors: 
\href{mailto:siyuyi@scu.edu.cn}{siyuyi@scu.edu.cn}, 
\href{mailto:juwei@scu.edu.cn}{juwei@scu.edu.cn}. 
\quad\quad
\textdagger\ These authors contributed equally.}

\abstract{
\textbf{Motivation:} Gene regulatory network inference from single-cell RNA sequencing (scRNA-seq) data is important for uncovering cell-state-specific transcriptional programs. However, scRNA-seq measurements are sparse and noisy, and experimentally validated TF--target interactions remain limited, making reliable inference challenging. Although graph neural networks have advanced GRN prediction, existing methods often rely on biologically unconstrained graph augmentation, such as random edge perturbation, and insufficiently control information transfer between genes and cells. These limitations may distort regulatory structures and weaken robustness under noisy and weakly supervised settings.
\\\textbf{Results:} To address these issues, we propose an innovative framework named Biological Evidence Refinement and Heterogeneous Dynamic Gating for Gene Regulatory Networks (BRIDGE). BRIDGE extracts gene and cell representations from the expression matrix and its matrix dual, and performs contrastive learning in the gene space and cell space between self and neighbors across the co-expression-refined regulatory view and the original graph. It then applies heterogeneous gated encoding to adaptively regulate information transfer between genes and cells, enabling robust transcription factor-to-target gene prediction. Experiments on benchmark datasets spanning three network types and seven cell types show that BRIDGE achieves state-of-the-art AUROC and AUPRC in most settings. In particular, on Specific networks, BRIDGE improves average AUPRC by 5\% over the second-best baseline, GCLink. In cross-cell-type few-shot transfer, BRIDGE consistently outperforms GCLink and GENELink across all six target cell types. A case study on hESC further supports the biological relevance of the predictions, with 9 of the top 10 and 46 of the top 100 novel TF--target interactions validated by ChIPBase.
\\\textbf{Availability:} The datasets and source code used in this study are publicly available online at \url{https://github.com/ShanwenTan/BRIDGE}.
}

\newskip\origskipfootins
\origskipfootins=\skip\footins
\skip\footins=0pt
\maketitle
\thispagestyle{empty}
\skip\footins=\origskipfootins
\setcounter{footnote}{0}

\section{Introduction}

Gene regulatory networks (GRNs) describe regulatory interactions between transcription factors (TFs) and target genes that govern transcription and cellular behavior \citep{marbach2012wisdom, pratapa2020benchmarking}. Recent advances in high-throughput sequencing, especially scRNA-seq, have enabled gene expression profiling at single-cell resolution \citep{kolodziejczyk2015technology}. With scRNA-seq data, GRN reconstruction can reveal cell-type-specific regulatory programs and clarify how gene interactions drive diverse cellular states, thereby supporting studies of cellular heterogeneity, disease mechanisms, and therapeutic development \citep{chan2017gene}. However, GRN inference from scRNA-seq remains challenging because single-cell measurements are inherently sparse and noisy, while experimentally validated regulatory interactions are often limited \citep{wang2023inferring, kommu2025prediction}. Under such sparsity and variability at the single-cell level, solely relying on gene embeddings, without explicitly modeling cell-state heterogeneity, can obscure rare or transient cellular states and consequently blur state-specific regulatory programs, making robust inference under weak supervision particularly difficult.

Early GRN inference methods mainly relied on information-theoretic measures and classical machine learning to recover regulatory dependencies from coexpression patterns and expression variability \citep{ margolin2006aracne, tsai2020grema, chan2017gene, matsumoto2017scode}. More recently, supervised deep learning and graph-based approaches have become increasingly important for GRN prediction from scRNA-seq data. For example, CNNC and DeepDRIM learn regulatory relationships from TF--target expression patterns using deep neural architectures \citep{yuan2019deep, chen2021deepdrim}. scGREAT further incorporates transformer-based gene representations derived from scRNA-seq data and biological text knowledge \citep{wang2024scgreat}, while GENELink formulates GRN inference as graph-based link prediction with graph attention \citep{chen2022graph}. DeepRIG improves robustness by constructing a prior coexpression graph and learning gene embeddings through a graph autoencoder \citep{wang2023inferring}. HGATLink moves beyond gene-only modeling by jointly learning gene and cell representations in a heterogeneous graph \citep{sun2025hgatlink}. Despite these advances, existing methods still face two major limitations: many remain largely gene-centric and therefore do not explicitly preserve cell-state heterogeneity, while heterogeneous graph methods often lack fine-grained control over cross-type information propagation between genes and cells, making them vulnerable to noise under weak supervision. These limitations become especially pronounced when supervision is sparse and biological priors are noisy or incomplete.

Contrastive learning has recently emerged as an effective strategy for improving graph representation robustness \citep{you2020graph, chen2020simple}. In the GRN setting, GCLink introduces graph contrastive learning to improve regulatory prediction performance \citep{yu2025gclink}. More broadly, methods such as AD-GCL, SimGRACE, and AFGRL explore adaptive perturbation or augmentation-free contrastive learning for general graphs \citep{suresh2021adversarial, xia2022simgrace, lee2022augmentation}. However, these strategies are mostly designed for generic graph learning and may either perturb biologically plausible edges or neglect structural constraints that are critical for regulatory graphs. This issue is particularly problematic in non-specific networks and STRING priors, where evidence is incomplete and noise is prevalent. Therefore, the core challenge is not simply to improve link prediction accuracy, but to achieve robust GRN inference from sparse and noisy scRNA-seq data by jointly preserving cell-state heterogeneity, constraining biologically implausible perturbations, and controlling cross-type information propagation between genes and cells. However, these requirements are still not jointly addressed by existing GRN inference frameworks.

Motivated by these limitations, we propose Biological Evidence Refinement and Heterogeneous Dynamic Gating for gene Regulatory networks (BRIDGE), a unified framework for robust GRN inference from scRNA-seq data. BRIDGE proceeds in three stages. First, Biological Evidence Refinement constructs evidence-guided graph views by pruning weakly supported regulatory edges, enabling biologically grounded view augmentation for contrastive learning. Second, Heterogeneous Dynamic Gated Representation Learning builds a heterogeneous graph over genes and cells and adaptively regulates cross-type message passing to reduce noise propagation while preserving cell-state information. Third, Dual-Space Neighborhood Contrastive Learning aligns gene and cell representations across views with neighborhood-aware multi-positive supervision, and a cell-conditioned decoder scores candidate TF--target regulations under cellular context. Through this design, BRIDGE addresses the key unmet need for a robust framework that integrates biologically grounded augmentation, explicit gene--cell heterogeneity modeling, and controlled heterogeneous information flow for reliable GRN inference. Extensive experiments on benchmark datasets spanning three types of real regulatory networks and seven cell types demonstrate that BRIDGE consistently outperforms strong baselines and achieves robust generalization, with particularly clear advantages under noisy and weakly supervised settings.
\begin{figure*}[t]
  \centering
  \includegraphics[width=\textwidth]{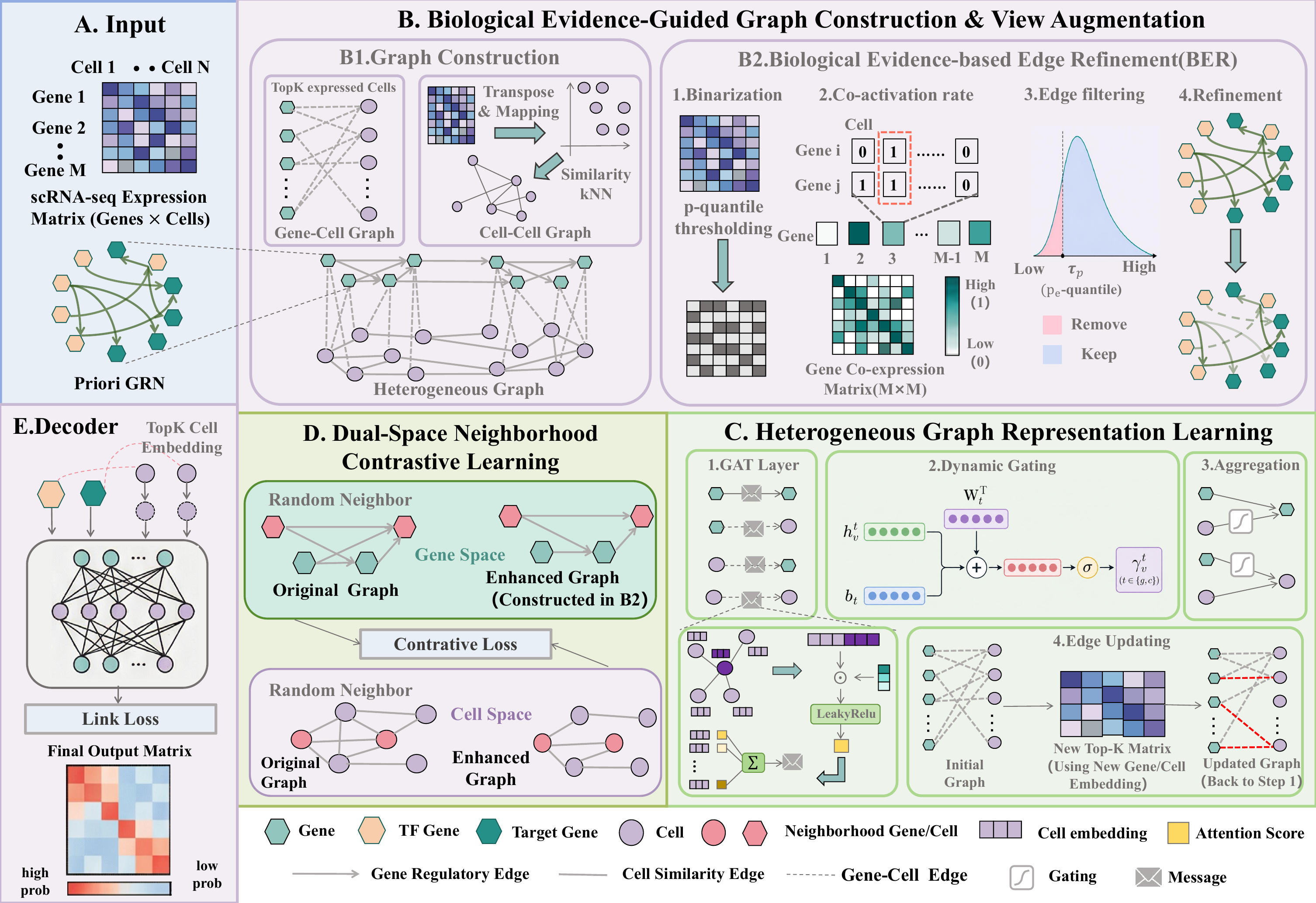}
  \caption{Overall framework of BRIDGE. Given a single-cell sequencing expression matrix and partial GRN regulatory relationships, BRIDGE first constructs an enhanced view based on biological prior information and cell-to-cell relationships built using $k$NN.  Then, it performs joint representation learning of genes and cells in a heterogeneous graph through a gating mechanism. Finally, during the decoding process, BRIDGE outputs the probability scores of different gene regulatory pairs under specific cellular conditions.}
  \label{fig:srta_framework}
\end{figure*}
\section{Materials and methods}
\label{sec:method2}
In this section, we describe the proposed method in detail. We first present the problem formulation, followed by the model components and the training objective. Finally, we introduce the datasets used in our experiments. Figure~\ref{fig:srta_framework} provides an overview of the BRIDGE framework.A full pseudocode listing is provided in Appendix~\ref{app:alg}.
\subsection{Problem Formulation}
We model a gene regulatory network as a directed graph over genes with node set $\mathcal{V}$, where $\mathcal{T} \subseteq \mathcal{V}_g$ denotes the subset of transcription factors (TFs). 
Given scRNA-seq measurements $\mathbf{X} \in \mathbb{R}^{G \times C}$ from $C$ cells and $G$ genes, we construct a heterogeneous graph 
$\mathcal{G} = (\mathcal{V}_g,\mathcal{V}_c, {A}^{gg}, A^{cc}, A^{gc}, A^{cg})$, where $\mathcal{V}_g$ and $\mathcal{V}_c$ denote the sets of gene and cell nodes, respectively, where $A^{gg}, A^{cc}, A^{gc}, A^{cg}$ denote gene--gene, cell--cell, gene--cell, and cell--gene adjacencies, respectively.

Specifically, ${A}^{gg} \in \{0,1\}^{G \times G}$ encodes gene--gene regulatory interactions, where $A_{i,j} = 1$ indicates a validated regulation $i \rightarrow j$. Our goal is to predict missing regulatory interactions by learning a scoring function over the heterogeneous graph:
\begin{equation}
\hat{A}_{i,j} = \sigma\!\left(f_{\theta}\!\left(i, j \,; \mathcal{G}\right)\right), 
\quad i \in \mathcal{T},\; j \in \mathcal{V}_g.
\end{equation}
\subsection{Biological Evidence-Guided Graph Construction and View Augmentation}
\label{subsec:BER}

\par\medskip
\noindent\textbf{Graph Construction}
\par
To jointly model gene–gene and gene–cell interactions, we construct a heterogeneous graph consisting of a gene graph, a cell graph, and a gene–cell bipartite graph. Given the gene expression matrix $X \in \mathbb{R}^{G \times C}$, where $G$ and $C$ denote the numbers of genes and cells respectively, the cell features are naturally defined as the transpose of gene features, i.e., $X^{c} = X^{\top}$.

The cell–cell adjacency matrix $A^{cc}$ is constructed based on a $k$-nearest neighbor (\textit{kNN}) graph over the cell feature space $X^{c}$. Specifically, we first define a symmetric neighborhood structure:
\begin{equation}
A^{cc}_{uv}
=
\frac{
\mathbb{I}\!\left(v \in \mathcal{N}_k(u; X^{c})\ \text{or}\ u \in \mathcal{N}_k(v; X^{c})\right)
}{
\sum\limits_{v'}
\mathbb{I}\!\left(v' \in \mathcal{N}_k(u; X^{c})\ \text{or}\ u \in \mathcal{N}_k(v'; X^{c})\right)
}
.
\label{eq:acc_knn}
\end{equation}
where $\mathcal{N}_k(u; X^c)$ denotes the set of $k$ nearest neighbors of cell $u$ in the feature space $X^c$, using expression similarity.

To capture cross-type interactions, we construct a gene–cell bipartite graph by selecting the top-$k$ cells with the highest expression for each gene:
\begin{equation}
A^{gc}_{iu}
=
\mathbb{I}(u \in \mathcal{K}_i),
\qquad
A^{cg} = (A^{gc})^{\top}.
\label{eq:agc_topk}
\end{equation}
where $\mathcal{K}_i$ denotes the set of the top-$k$ cells with the highest expression values for gene $i$.
This construction enables the model to propagate information across gene and cell domains, forming the basis for heterogeneous message passing.
\par\medskip
\noindent\textbf{Biological Evidence-based Edge Refinement (BER)}
\par
\begin{figure}[t]
  \centering
  \includegraphics[width=\columnwidth]{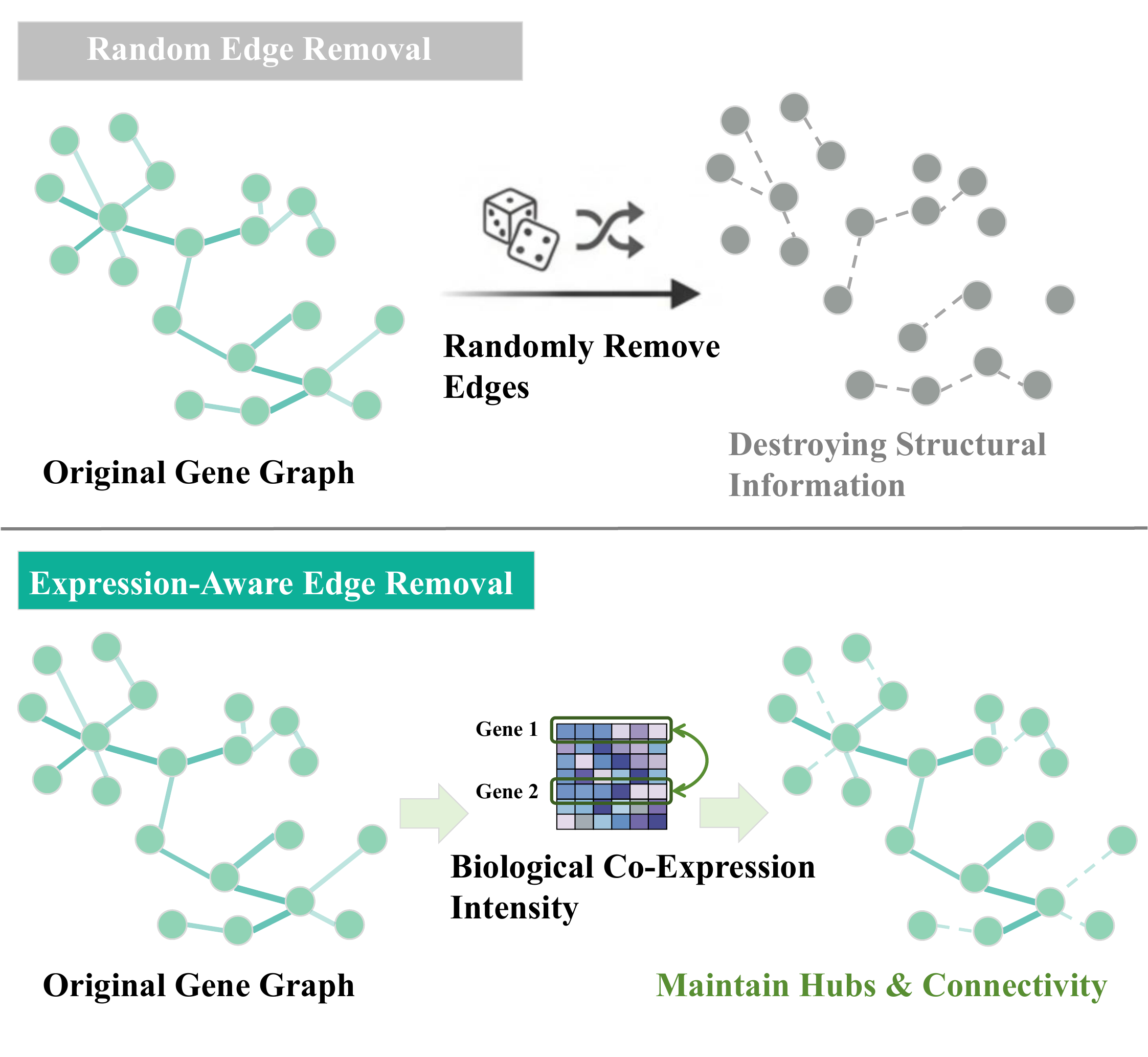}
  \caption{Comparison of edge-removal augmentations. (a) Random edge dropping removes links indiscriminately, disrupting key topology and important structural motifs. (b) Expression-aware removal uses scRNA-seq co-expression evidence to prune low-support edges, effectively denoising the graph while preserving high-confidence hubs and overall connectivity.}
  \label{fig:ber_aug}
\end{figure}

The graph augmentation procedure is illustrated in Figure~\ref{fig:ber_aug}. Specifically, we leverage gene co-activation patterns in single-cell expression data to perform biologically evidence-guided structural refinement on the gene--gene relationships of the original heterogeneous graph for view augmentation.

Formally, for any gene pair $(i,j)$, we first define their co-activation strength as
\begin{equation}
w_{ij}
=
\frac{1}{C}
\sum_{u=1}^{C}
\mathbb{I}(X_{iu} > \tau)\,\mathbb{I}(X_{ju} > \tau).
\end{equation}
where $\tau$ is determined as the value corresponding to the $\rho$-quantile of all entries in the expression matrix $X$, and is used to identify whether a gene is highly expressed in a given cell. This measure captures the frequency with which genes $i$ and $j$ are simultaneously highly expressed across the cell population, and thus serves as a proxy for the biological support of the regulatory edge $i \rightarrow j$.

Based on the distribution of co-activation strengths over all observed gene--gene edges, we further filter out edges with weak biological evidence. Let $\tau_p$ denote the $p_e$-quantile of these co-activation scores. The augmented gene--gene adjacency matrix is then defined as
\begin{equation}
\tilde{A}^{gg}_{ij}
=
A^{gg}_{ij} \cdot \mathbb{I}(w_{ij} > \tau_p).
\label{eq:aug_refine}
\end{equation}
Accordingly, we construct two heterogeneous graph views for contrastive learning, namely the original graph $\mathcal{G}$ and its biologically refined augmented counterpart $\tilde{\mathcal{G}}$.

\subsection{Heterogeneous Dynamic Gated Representation Learning}

To jointly model the heterogeneous relationships between genes and cells, we introduce a relation-aware graph attention mechanism over the constructed heterogeneous graph, enabling message passing across different types of edges. Specifically, for each relation type $\phi \in \{gg, gc, cc, cg\}$, we define a relation-specific graph attention aggregation over the corresponding neighborhood $\mathcal{N}_{\phi}(v)$, yielding the message representation of node $v$ under relation $\phi$:
\begin{equation}
\mathbf{m}_{v}^{\phi}
=
\operatorname{GAT}_{\phi}
\bigl(
\mathbf{h}_{v},
\mathbf{h}_{u},\ u\!\in\!\mathcal{N}_{\phi}(v)
\bigr).
\label{eq:gat_message}
\end{equation}
The detailed formulation of $\operatorname{GAT}_{\phi}$, including attention weight computation and normalization across all relation types $\phi \in \{gg, gc, cc, cg\}$, is provided in Appendix~\ref{app:hgat}.

Building upon this, we further introduce a dynamic gating mechanism to adaptively fuse intra-type and cross-type information. For gene nodes, the updated representation is determined by both gene--gene and gene--cell messages:
First, we introduce dynamic gating coefficients to adaptively control the contribution of cross-type information. For gene and cell nodes, the gating coefficients are defined as
\begin{equation}
\gamma_v^{t}
=
\sigma\!\left(
\mathbf{w}_{t}^\top \mathbf{h}_v^{t} + b_{t}
\right),
\qquad
t \in \{g,c\}.
\label{eq:gating_coef}
\end{equation}
Based on the learned gating coefficients, we fuse intra-type and cross-type messages to update node representations. Specifically, for gene nodes, the updated representation is given by
\begin{equation}
\mathbf{h}_i^{g'}
=
\operatorname{LeakyReLU}
\left(
\mathbf{h}_i^g
+
\mathbf{m}_i^{gg}
+
\gamma_i^g \mathbf{m}_i^{gc}
\right).
\label{eq:update_gene}
\end{equation}
A similar update rule is applied to cell nodes, where intra-cell and cross-type messages are combined via gating. This mechanism regulates cross-type information flow in a node-specific manner, reducing noise propagation and improving representation robustness. 

Furthermore, to better capture heterogeneous structures, we incorporate a dynamic graph reconstruction strategy during encoding. After the first layer, gene–cell connections are updated by selecting, for each gene, the most similar cells based on cosine similarity of embeddings. This process progressively refines cross-type interactions and improves alignment between gene and cell representations. The details of this module are in Appendix~\ref{app:hgat}.
\subsection{Dual-Space Neighborhood Contrastive Learning}

To enhance representation consistency and robustness under graph structural perturbations, we construct a dual-view contrastive learning framework based on the original graph $\mathcal{G}$ and its biologically refined counterpart $\tilde{\mathcal{G}}$. Specifically, heterogeneous graph encoding is performed on both $\mathcal{G}$ and $\tilde{\mathcal{G}}$ to obtain node representations $\mathbf{h}_i^{\mathcal{G}}$ and $\mathbf{h}_i^{\tilde{\mathcal{G}}}$, and cross-view alignment is enforced between them.

Unlike conventional contrastive learning that considers only identical nodes as positive pairs, we incorporate graph structural information by introducing a neighborhood-aware multi-positive strategy. For each node $i$, the positive set is defined as $\mathcal{P}(i)=\{i, j(i)\}$, where $j(i)$ denotes a randomly sampled neighbor. We measure similarity using cosine similarity:
\begin{equation}
\mathrm{sim}(\mathbf{a},\mathbf{b})
=
\frac{\mathbf{a}^\top \mathbf{b}}{\|\mathbf{a}\|_2\|\mathbf{b}\|_2}.
\end{equation}

Based on this similarity, the contrastive objective is formulated as:
\begin{equation}
\mathcal{L}_{\mathrm{con}}
=
-\frac{1}{N}\sum_{i=1}^{N}
\log
\frac{
\sum_{k\in\mathcal{P}(i)}
\exp\!\left(\frac{\mathrm{sim}(\mathbf{h}_i^{\mathcal{G}}, \mathbf{h}_k^{\tilde{\mathcal{G}}})}{\tau}\right)
}{
\sum_{k}
\exp\!\left(\frac{\mathrm{sim}(\mathbf{h}_i^{\mathcal{G}}, \mathbf{h}_k^{\tilde{\mathcal{G}}})}{\tau}\right)
}.
\label{eq:con_loss}
\end{equation}

Furthermore, the above contrastive learning procedure is independently applied in both the gene space and the cell space, yielding $\mathcal{L}_{\mathrm{con}}^{g}$ and $\mathcal{L}_{\mathrm{con}}^{c}$. This dual-space alignment simultaneously regularizes gene regulatory representations and cell state representations across views, thereby improving robustness to structural perturbations.
\subsection{Decoder and Link Prediction}

Given the gene representations $\mathbf{h}_i$ learned by the heterogeneous encoder, we further project them into task-specific embeddings to capture the distinct roles of transcription factors (TFs) and target genes in regulatory interactions, resulting in the TF-specific representation $\mathbf{z}_i^{\mathrm{tf}}$ and the target-specific representation $\mathbf{z}_i^{\mathrm{tar}}$.

To incorporate cell-state information, we construct a dynamic gene--cell context representation. Specifically, for each gene, we compute its similarity to all cell embeddings and select the top-$k$ most relevant cells for aggregation, yielding a context-aware representation that reflects the underlying cellular environment.

For a given TF--target pair $(t,r)$, we construct a joint representation by concatenating $\mathbf{z}_t^{\mathrm{tf}}$ and $\mathbf{z}_r^{\mathrm{tar}}$, along with their corresponding cell context features. The decoder maps this combined representation to a regulatory probability:
\begin{equation}
p_{tr}
=
\sigma\!\Big(
\mathbf{w}^\top
\big[
\mathbf{z}_t^{\mathrm{tf}} \,\Vert\,
\mathbf{z}_r^{\mathrm{tar}} \,\Vert\,
\mathbf{c}_t \,\Vert\,
\mathbf{c}_r
\big]
\Big).
\label{eq:decoder_score}
\end{equation}

During training, we adopt a dual-view supervision strategy, where predictions are performed on both the original graph $\mathcal{G}$ and the augmented graph $\tilde{\mathcal{G}}$. The link prediction loss is defined as:
\begin{equation}
\mathcal{L}_{\mathrm{link}}
=
\mathrm{BCE}\big(p_{tr}^{\mathcal{G}}, y_{tr}\big)
+
\mathrm{BCE}\big(p_{tr}^{\tilde{\mathcal{G}}}, y_{tr}\big).
\label{eq:link_loss}
\end{equation}

The final training objective combines the link prediction loss with the dual-space contrastive learning objectives in both gene and cell spaces through weighted summation, enabling the model to achieve accurate regulatory prediction while maintaining representation consistency across graph perturbations.
\subsection{Datasets}
We follow the same setting as BEELINE \citep{pratapa2020benchmarking}, adopting benchmark scRNA-seq datasets spanning seven cell types: five mouse cell types (mESC, mHSC-E, mHSC-GM, mHSC-L, and mDC) and two human cell types (hESC and hHEP). We use three real regulatory networks from different sources as supervisory signals: (i) Cell-type-specific \citep{encode2020expanded} chromatin immunoprecipitation sequencing (ChIP-seq) networks provide high-quality supervisory information; (ii) Non-specific ChIP-seq networks \citep{garcia2019benchmark} are used to evaluate BRIDGE's generalization ability across different cell types; (iii) STRING functional interaction network \citep{szklarczyk2019string} was included to provide weak supervisory signals for transcription factor-target gene prediction.

Specifically, following the same preprocessing protocol as GCLink~\citep{yu2025gclink} and BEELINE~\citep{pratapa2020benchmarking}, we first remove genes expressed in fewer than $10\%$ of cells. We then apply Bonferroni correction and discard genes with adjusted $p$-values greater than $0.01$. Finally, we select the top $500$ and $1000$ highly variable genes based on the variance ranking strategy proposed in BEELINE. The detailed dataset statistics are reported in Appendix~\ref{app:datasetstats} (Table~\ref{tab:dataset_statistics}).

For dataset splitting, we adopt tailored strategies according to network characteristics. For the Specific networks, we perform \emph{per-transcription-factor (TF) synchronized splitting}, where positive TF--target pairs are partitioned into training, validation, and test sets using fixed proportions ($2/3$ for training, $1/5$ for validation, and the remainder for testing). All non-target genes are treated as \emph{hard negatives} and are partitioned consistently with the corresponding positive samples to preserve TF-level structural properties.

For the Non-specific and STRING networks, positive samples are first split into training and test sets with a ratio of $2/3$ and $1/3$, respectively. One-fifth of the training positives are further held out as a validation set. Negative samples are randomly sampled at a $1\!:\!1$ ratio with positive samples for training and validation, while the number of negative samples in the test set is determined according to the true network density to better reflect real-world sparsity. To mitigate randomness, the sampling process is repeated five times, and the average performance is reported.

\section{Experiment}
In this investigation, we conduct a comprehensive evaluation of BRIDGE across various cell types and regulatory networks. We examine its performance in inferring gene regulatory relationships, its generalization ability across cell types, the effectiveness of its key components, its robustness to noisy regulatory signals, and its stability under different hyperparameter settings.

\begin{table*}[t]
\centering
\scriptsize
\fontfamily{ptm}\selectfont
\setlength{\tabcolsep}{2.0pt}
\renewcommand{\arraystretch}{1.10}

\definecolor{ZebraGray}{gray}{0.95}
\definecolor{TFRowGray}{gray}{0.85}
\definecolor{darkpurple}{rgb}{0.4,0,0.4} 

\captionof{table}{GRN regulation inference. We show the AUPRC and AUROC metrics for predicting gene regulatory relationships under different network types and cell types. \textbf{Bold} numbers indicate the optimal values, and \underline{underlined} numbers indicate the second best values.}
\begin{adjustbox}{max width=\textwidth}
\begin{tabular}{ll
    >{\columncolor{ZebraGray}}c c
    >{\columncolor{ZebraGray}}c c
    >{\columncolor{ZebraGray}}c c
    >{\columncolor{ZebraGray}}c c
    >{\columncolor{ZebraGray}}c c
    >{\columncolor{ZebraGray}}c c
    >{\columncolor{ZebraGray}}c c
    >{\columncolor{ZebraGray}}c c
}
\toprule
\rowcolor{HeadPurple}
Network Type & Cell Type &
\multicolumn{2}{c}{GENELink} &
\multicolumn{2}{c}{GMFGRN} &
\multicolumn{2}{c}{GCLink} &
\multicolumn{2}{c}{scTransNet} &
\multicolumn{2}{c}{HGATLink} &
\multicolumn{2}{c}{CNNC} &
\multicolumn{2}{c}{GNE} &
\multicolumn{2}{c}{\textbf{BRIDGE}} \\
\cmidrule(lr){3-4}\cmidrule(lr){5-6}\cmidrule(lr){7-8}\cmidrule(lr){9-10}\cmidrule(lr){11-12}\cmidrule(lr){13-14}\cmidrule(lr){15-16}\cmidrule(lr){17-18}
\rowcolor{HeadPurple}
& &
AUPRC & AUROC &
AUPRC & AUROC &
AUPRC & AUROC &
AUPRC & AUROC &
AUPRC & AUROC &
AUPRC & AUROC &
AUPRC & AUROC &
AUPRC & AUROC \\
\midrule
\rowcolor{TFRowGray}
\multicolumn{18}{c}{\textbf{TF+1000}} \\
\midrule

\multirow{7}{*}{Specific}
& hESC
& 0.50\color{darkpurple}\tiny{±0.004} & 0.82\color{darkpurple}\tiny{±0.001}
& 0.59\color{darkpurple}\tiny{±0.002} & 0.86\color{darkpurple}\tiny{±0.005}
& 0.55\color{darkpurple}\tiny{±0.005} & 0.86\color{darkpurple}\tiny{±0.002}
& \underline{0.60}\color{darkpurple}\tiny{±0.001} & \underline{0.87}\color{darkpurple}\tiny{±0.004}
& 0.60\color{darkpurple}\tiny{±0.003} & 0.87\color{darkpurple}\tiny{±0.003}
& 0.42\color{darkpurple}\tiny{±0.005} & 0.58\color{darkpurple}\tiny{±0.003}
& 0.33\color{darkpurple}\tiny{±0.002} & 0.54\color{darkpurple}\tiny{±0.004}
& \textbf{0.61}\color{darkpurple}\tiny{±0.004} & \textbf{0.88}\color{darkpurple}\tiny{±0.001} \\

& hHEP
& 0.69\color{darkpurple}\tiny{±0.001} & 0.82\color{darkpurple}\tiny{±0.004}
& 0.81\color{darkpurple}\tiny{±0.004} & 0.81\color{darkpurple}\tiny{±0.001}
& 0.78\color{darkpurple}\tiny{±0.003} & 0.87\color{darkpurple}\tiny{±0.003}
& 0.82\color{darkpurple}\tiny{±0.002} & 0.85\color{darkpurple}\tiny{±0.005}
& \underline{0.82}\color{darkpurple}\tiny{±0.003} & \underline{0.89}\color{darkpurple}\tiny{±0.003}
& 0.64\color{darkpurple}\tiny{±0.005} & 0.68\color{darkpurple}\tiny{±0.002}
& 0.54\color{darkpurple}\tiny{±0.001} & 0.72\color{darkpurple}\tiny{±0.004}
& \textbf{0.83}\color{darkpurple}\tiny{±0.002} & \textbf{0.90}\color{darkpurple}\tiny{±0.003} \\

& mDC
& 0.11\color{darkpurple}\tiny{±0.003} & 0.73\color{darkpurple}\tiny{±0.003}
& 0.12\color{darkpurple}\tiny{±0.001} & 0.74\color{darkpurple}\tiny{±0.004}
& \underline{0.13}\color{darkpurple}\tiny{±0.005} & \underline{0.75}\color{darkpurple}\tiny{±0.002}
& \underline{0.12}\color{darkpurple}\tiny{±0.002} & 0.75\color{darkpurple}\tiny{±0.001}
& 0.12\color{darkpurple}\tiny{±0.004} & 0.73\color{darkpurple}\tiny{±0.003}
& 0.05\color{darkpurple}\tiny{±0.003} & 0.67\color{darkpurple}\tiny{±0.005}
& 0.08\color{darkpurple}\tiny{±0.003} & 0.62\color{darkpurple}\tiny{±0.002}
& \textbf{0.14}\color{darkpurple}\tiny{±0.004} & \textbf{0.76}\color{darkpurple}\tiny{±0.001} \\

& mESC
& 0.75\color{darkpurple}\tiny{±0.002} & 0.85\color{darkpurple}\tiny{±0.004}
& 0.82\color{darkpurple}\tiny{±0.005} & 0.90\color{darkpurple}\tiny{±0.001}
& 0.78\color{darkpurple}\tiny{±0.001} & 0.90\color{darkpurple}\tiny{±0.003}
& \underline{0.84}\color{darkpurple}\tiny{±0.004} & \underline{0.92}\color{darkpurple}\tiny{±0.003}
& 0.83\color{darkpurple}\tiny{±0.002} & 0.91\color{darkpurple}\tiny{±0.005}
& 0.74\color{darkpurple}\tiny{±0.003} & 0.62\color{darkpurple}\tiny{±0.001}
& 0.47\color{darkpurple}\tiny{±0.003} & 0.72\color{darkpurple}\tiny{±0.004}
& \textbf{0.86}\color{darkpurple}\tiny{±0.003} & \textbf{0.93}\color{darkpurple}\tiny{±0.002} \\

& mHSC-E
& 0.89\color{darkpurple}\tiny{±0.004} & 0.82\color{darkpurple}\tiny{±0.002}
& \underline{0.94}\color{darkpurple}\tiny{±0.001} & 0.90\color{darkpurple}\tiny{±0.005}
& 0.92\color{darkpurple}\tiny{±0.005} & 0.87\color{darkpurple}\tiny{±0.012}
& \underline{0.94}\color{darkpurple}\tiny{±0.002} & 0.90\color{darkpurple}\tiny{±0.004}
& \underline{0.94}\color{darkpurple}\tiny{±0.003} & \underline{0.91}\color{darkpurple}\tiny{±0.001}
& 0.90\color{darkpurple}\tiny{±0.004} & 0.53\color{darkpurple}\tiny{±0.003}
& 0.80\color{darkpurple}\tiny{±0.008} & 0.61\color{darkpurple}\tiny{±0.005}
& \textbf{0.95}\color{darkpurple}\tiny{±0.002} & \textbf{0.94}\color{darkpurple}\tiny{±0.004} \\

& mHSC-GM
& 0.90\color{darkpurple}\tiny{±0.005} & 0.89\color{darkpurple}\tiny{±0.002}
& \underline{0.94}\color{darkpurple}\tiny{±0.001} & 0.91\color{darkpurple}\tiny{±0.004}
& 0.92\color{darkpurple}\tiny{±0.003} & 0.90\color{darkpurple}\tiny{±0.001}
& \underline{0.94}\color{darkpurple}\tiny{±0.003} & 0.90\color{darkpurple}\tiny{±0.005}
& \textbf{0.95}\color{darkpurple}\tiny{±0.004} & \underline{0.92}\color{darkpurple}\tiny{±0.003}
& 0.88\color{darkpurple}\tiny{±0.002} & 0.59\color{darkpurple}\tiny{±0.003}
& 0.79\color{darkpurple}\tiny{±0.005} & 0.62\color{darkpurple}\tiny{±0.004}
& \textbf{0.95}\color{darkpurple}\tiny{±0.001} & \textbf{0.93}\color{darkpurple}\tiny{±0.002} \\

& mHSC-L
& 0.81\color{darkpurple}\tiny{±0.002} & 0.81\color{darkpurple}\tiny{±0.004}
& 0.82\color{darkpurple}\tiny{±0.005} & 0.82\color{darkpurple}\tiny{±0.001}
& 0.83\color{darkpurple}\tiny{±0.001} & 0.82\color{darkpurple}\tiny{±0.003}
& \underline{0.84}\color{darkpurple}\tiny{±0.004} & \underline{0.86}\color{darkpurple}\tiny{±0.003}
& 0.84\color{darkpurple}\tiny{±0.003} & 0.85\color{darkpurple}\tiny{±0.005}
& 0.74\color{darkpurple}\tiny{±0.003} & 0.62\color{darkpurple}\tiny{±0.002}
& 0.66\color{darkpurple}\tiny{±0.004} & 0.56\color{darkpurple}\tiny{±0.001}
& \textbf{0.87}\color{darkpurple}\tiny{±0.002} & \textbf{0.87}\color{darkpurple}\tiny{±0.003} \\

\hline

\multirow{7}{*}{Non-Specific}
& hESC
& \underline{0.03}\color{darkpurple}\tiny{±0.009} & 0.64\color{darkpurple}\tiny{±0.008}
& \textbf{0.04}\color{darkpurple}\tiny{±0.010} & \underline{0.67}\color{darkpurple}\tiny{±0.009}
& \textbf{0.04}\color{darkpurple}\tiny{±0.008} & \underline{0.67}\color{darkpurple}\tiny{±0.010}
& \textbf{0.04}\color{darkpurple}\tiny{±0.011} & 0.67\color{darkpurple}\tiny{±0.007}
& \textbf{0.04}\color{darkpurple}\tiny{±0.005} & \underline{0.67}\color{darkpurple}\tiny{±0.006}
& 0.02\color{darkpurple}\tiny{±0.010} & 0.51\color{darkpurple}\tiny{±0.011}
& 0.02\color{darkpurple}\tiny{±0.009} & 0.55\color{darkpurple}\tiny{±0.010}
& \textbf{0.04}\color{darkpurple}\tiny{±0.006} & \textbf{0.69}\color{darkpurple}\tiny{±0.006} \\

& hHEP
& \textbf{0.05}\color{darkpurple}\tiny{±0.010} & \textbf{0.69}\color{darkpurple}\tiny{±0.033}
& \underline{0.04}\color{darkpurple}\tiny{±0.009} & \textbf{0.69}\color{darkpurple}\tiny{±0.031}
& \underline{0.04}\color{darkpurple}\tiny{±0.011} & \underline{0.68} \color{darkpurple}\tiny{±0.036}
& \underline{0.04}\color{darkpurple}\tiny{±0.012} & 0.67\color{darkpurple}\tiny{±0.028}
& \underline{0.04}\color{darkpurple}\tiny{±0.007} & \underline{0.68}\color{darkpurple}\tiny{±0.029}
& 0.01\color{darkpurple}\tiny{±0.012} & 0.53\color{darkpurple}\tiny{±0.037}
& 0.02\color{darkpurple}\tiny{±0.010} & 0.54\color{darkpurple}\tiny{±0.034}
& \underline{0.04}\color{darkpurple}\tiny{±0.008} & \textbf{0.69} \color{darkpurple}\tiny{±0.030} \\

& mDC
& 0.06\color{darkpurple}\tiny{±0.013} & 0.73\color{darkpurple}\tiny{±0.009}
& 0.12\color{darkpurple}\tiny{±0.012} & 0.73\color{darkpurple}\tiny{±0.010}
& 0.11\color{darkpurple}\tiny{±0.011} & 0.73\color{darkpurple}\tiny{±0.011}
& \underline{0.15}\color{darkpurple}\tiny{±0.010} & \underline{0.74}\color{darkpurple}\tiny{±0.008}
& 0.12\color{darkpurple}\tiny{±0.007} & \underline0.74\color{darkpurple}\tiny{±0.006}
& 0.02\color{darkpurple}\tiny{±0.012} & 0.63\color{darkpurple}\tiny{±0.012}
& 0.02\color{darkpurple}\tiny{±0.010} & 0.54\color{darkpurple}\tiny{±0.011}
& \textbf{0.15}\color{darkpurple}\tiny{±0.008} & \textbf{0.77}\color{darkpurple}\tiny{±0.006} \\

& mESC
& 0.04\color{darkpurple}\tiny{±0.022} & 0.73\color{darkpurple}\tiny{±0.030}
& \underline{0.05} \color{darkpurple}\tiny{±0.021} & 0.73\color{darkpurple}\tiny{±0.031}
& \textbf{0.07}\color{darkpurple}\tiny{±0.020} & \textbf{0.75}\color{darkpurple}\tiny{±0.029}
& \underline{0.05} \color{darkpurple}\tiny{±0.018} & \underline{0.75}\color{darkpurple}\tiny{±0.024}
& 0.04\color{darkpurple}\tiny{±0.015} & 0.72\color{darkpurple}\tiny{±0.026}
& 0.02\color{darkpurple}\tiny{±0.023} & 0.53\color{darkpurple}\tiny{±0.032}
& 0.02\color{darkpurple}\tiny{±0.020} & 0.56\color{darkpurple}\tiny{±0.030}
& \underline{0.05}\color{darkpurple}\tiny{±0.016} & \textbf{0.75}\color{darkpurple}\tiny{±0.027} \\

& mHSC-E
& 0.10\color{darkpurple}\tiny{±0.020} & 0.74\color{darkpurple}\tiny{±0.016}
& 0.14\color{darkpurple}\tiny{±0.018} & 0.74\color{darkpurple}\tiny{±0.017}
& \underline{0.15}\color{darkpurple}\tiny{±0.019} & \underline{0.75}\color{darkpurple}\tiny{±0.018}
& \textbf{0.18}\color{darkpurple}\tiny{±0.017} & \underline{0.75}\color{darkpurple}\tiny{±0.012}
& \underline{0.15}\color{darkpurple}\tiny{±0.012} & \underline{0.75}\color{darkpurple}\tiny{±0.009}
& 0.02\color{darkpurple}\tiny{±0.021} & 0.65\color{darkpurple}\tiny{±0.019}
& 0.03\color{darkpurple}\tiny{±0.018} & 0.59\color{darkpurple}\tiny{±0.017}
& \textbf{0.18}\color{darkpurple}\tiny{±0.015} & \textbf{0.76}\color{darkpurple}\tiny{±0.010} \\

& mHSC-GM
& 0.24\color{darkpurple}\tiny{±0.040} & 0.72\color{darkpurple}\tiny{±0.017}
& 0.21\color{darkpurple}\tiny{±0.039} & 0.75\color{darkpurple}\tiny{±0.018}
& 0.26\color{darkpurple}\tiny{±0.041} & 0.73\color{darkpurple}\tiny{±0.019}
& \underline{0.28}\color{darkpurple}\tiny{±0.033} & \underline{0.76}\color{darkpurple}\tiny{±0.012}
& 0.23\color{darkpurple}\tiny{±0.035} & 0.75\color{darkpurple}\tiny{±0.009}
& 0.03\color{darkpurple}\tiny{±0.040} & 0.67\color{darkpurple}\tiny{±0.020}
& 0.05\color{darkpurple}\tiny{±0.037} & 0.50\color{darkpurple}\tiny{±0.018}
& \textbf{0.29}\color{darkpurple}\tiny{±0.036} & \textbf{0.77}\color{darkpurple}\tiny{±0.010} \\

& mHSC-L
& 0.12\color{darkpurple}\tiny{±0.046} & \textbf{0.68}\color{darkpurple}\tiny{±0.035}
& 0.12\color{darkpurple}\tiny{±0.045} & \underline{0.67}\color{darkpurple}\tiny{±0.037}
& 0.12\color{darkpurple}\tiny{±0.048} & \underline{0.67}\color{darkpurple}\tiny{±0.038}
& \underline{0.16}\color{darkpurple}\tiny{±0.041} & \textbf{0.68}\color{darkpurple}\tiny{±0.031}
& 0.12\color{darkpurple}\tiny{±0.037} & 0.66\color{darkpurple}\tiny{±0.026}
& 0.05\color{darkpurple}\tiny{±0.049} & 0.66\color{darkpurple}\tiny{±0.039}
& 0.04\color{darkpurple}\tiny{±0.044} & 0.66\color{darkpurple}\tiny{±0.036}
& \textbf{0.17}\color{darkpurple}\tiny{±0.040} & \textbf{0.68}\color{darkpurple}\tiny{±0.029} \\

\hline

\multirow{7}{*}{STRING}
& hESC
& 0.16\color{darkpurple}\tiny{±0.018} & 0.79\color{darkpurple}\tiny{±0.015}
& 0.23\color{darkpurple}\tiny{±0.020} & 0.84\color{darkpurple}\tiny{±0.017}
& \textbf{0.25}\color{darkpurple}\tiny{±0.022} & \underline{0.86}\color{darkpurple}\tiny{±0.018}
& 0.24\color{darkpurple}\tiny{±0.016} & \underline{0.88}\color{darkpurple}\tiny{±0.012}
& 0.24\color{darkpurple}\tiny{±0.017} & 0.84\color{darkpurple}\tiny{±0.014}
& 0.04\color{darkpurple}\tiny{±0.023} & 0.52\color{darkpurple}\tiny{±0.028}
& 0.05\color{darkpurple}\tiny{±0.021} & 0.57\color{darkpurple}\tiny{±0.026}
& \textbf{0.25}\color{darkpurple}\tiny{±0.051} & \textbf{0.88}\color{darkpurple}\tiny{±0.007} \\

& hHEP
& 0.19\color{darkpurple}\tiny{±0.017} & 0.85\color{darkpurple}\tiny{±0.014}
& 0.23\color{darkpurple}\tiny{±0.019} & 0.86\color{darkpurple}\tiny{±0.016}
& \underline{0.24}\color{darkpurple}\tiny{±0.020} & \underline{0.87}\color{darkpurple}\tiny{±0.017}
& \underline{0.24}\color{darkpurple}\tiny{±0.015} & \underline{0.87}\color{darkpurple}\tiny{±0.011}
& \underline{0.24}\color{darkpurple}\tiny{±0.016} & 0.86\color{darkpurple}\tiny{±0.013}
& 0.03\color{darkpurple}\tiny{±0.024} & 0.59\color{darkpurple}\tiny{±0.030}
& 0.04\color{darkpurple}\tiny{±0.022} & 0.61\color{darkpurple}\tiny{±0.028}
& \textbf{0.25}\color{darkpurple}\tiny{±0.011} & \textbf{0.88}\color{darkpurple}\tiny{±0.006} \\

& mDC
& 0.26\color{darkpurple}\tiny{±0.026} & 0.86\color{darkpurple}\tiny{±0.021}
& 0.21\color{darkpurple}\tiny{±0.029} & 0.86\color{darkpurple}\tiny{±0.023}
& 0.24\color{darkpurple}\tiny{±0.028} & 0.87\color{darkpurple}\tiny{±0.024}
& 0.30\color{darkpurple}\tiny{±0.022} & \underline{0.88}\color{darkpurple}\tiny{±0.018}
& \underline{0.32}\color{darkpurple}\tiny{±0.024} & \underline{0.88}\color{darkpurple}\tiny{±0.019}
& 0.05\color{darkpurple}\tiny{±0.031} & 0.64\color{darkpurple}\tiny{±0.035}
& 0.22\color{darkpurple}\tiny{±0.034} & 0.62\color{darkpurple}\tiny{±0.032}
& \textbf{0.36}\color{darkpurple}\tiny{±0.008} & \textbf{0.89}\color{darkpurple}\tiny{±0.017} \\

& mESC
& 0.13\color{darkpurple}\tiny{±0.024} & 0.87\color{darkpurple}\tiny{±0.022}
& 0.20\color{darkpurple}\tiny{±0.026} & 0.88\color{darkpurple}\tiny{±0.024}
& \textbf{0.23}\color{darkpurple}\tiny{±0.025} & \underline{0.89}\color{darkpurple}\tiny{±0.023}
& 0.20\color{darkpurple}\tiny{±0.019} & \underline{0.89}\color{darkpurple}\tiny{±0.016}
& \underline{0.21}\color{darkpurple}\tiny{±0.021} & 0.88\color{darkpurple}\tiny{±0.018}
& 0.05\color{darkpurple}\tiny{±0.028} & 0.53\color{darkpurple}\tiny{±0.031}
& 0.04\color{darkpurple}\tiny{±0.026} & 0.55\color{darkpurple}\tiny{±0.029}
& \textbf{0.23}\color{darkpurple}\tiny{±0.009} & \textbf{0.90}\color{darkpurple}\tiny{±0.002} \\

& mHSC-E
& 0.19\color{darkpurple}\tiny{±0.036} & 0.85\color{darkpurple}\tiny{±0.028}
& 0.34\color{darkpurple}\tiny{±0.039} & 0.85\color{darkpurple}\tiny{±0.030}
& 0.26\color{darkpurple}\tiny{±0.035} & 0.85\color{darkpurple}\tiny{±0.029}
& \underline{0.35}\color{darkpurple}\tiny{±0.028} & \underline{0.86}\color{darkpurple}\tiny{±0.021}
& 0.34\color{darkpurple}\tiny{±0.031} & \underline{0.86}\color{darkpurple}\tiny{±0.024}
& 0.03\color{darkpurple}\tiny{±0.040} & 0.51\color{darkpurple}\tiny{±0.036}
& 0.06\color{darkpurple}\tiny{±0.037} & 0.58\color{darkpurple}\tiny{±0.033}
& \textbf{0.36}\color{darkpurple}\tiny{±0.033} & \textbf{0.87}\color{darkpurple}\tiny{±0.018} \\

& mHSC-GM
& 0.24\color{darkpurple}\tiny{±0.043} & 0.80\color{darkpurple}\tiny{±0.036}
& 0.38\color{darkpurple}\tiny{±0.046} & 0.82\color{darkpurple}\tiny{±0.038}
& \textbf{0.46}\color{darkpurple}\tiny{±0.044} & \underline{0.84}\color{darkpurple}\tiny{±0.041}
& \underline{0.43}\color{darkpurple}\tiny{±0.035} & \underline{0.84}\color{darkpurple}\tiny{±0.030}
& \underline{0.43}\color{darkpurple}\tiny{±0.039} & 0.82\color{darkpurple}\tiny{±0.033}
& 0.05\color{darkpurple}\tiny{±0.048} & 0.57\color{darkpurple}\tiny{±0.042}
& 0.08\color{darkpurple}\tiny{±0.045} & 0.57\color{darkpurple}\tiny{±0.040}
& \textbf{0.46}\color{darkpurple}\tiny{±0.032} & \textbf{0.87}\color{darkpurple}\tiny{±0.022} \\

& mHSC-L
& 0.27\color{darkpurple}\tiny{±0.050} & 0.73\color{darkpurple}\tiny{±0.041}
& 0.44\color{darkpurple}\tiny{±0.053} & 0.77\color{darkpurple}\tiny{±0.043}
& 0.35\color{darkpurple}\tiny{±0.055} & 0.75\color{darkpurple}\tiny{±0.044}
& \underline{0.45}\color{darkpurple}\tiny{±0.042} & \underline{0.79}\color{darkpurple}\tiny{±0.032}
& \underline{0.45}\color{darkpurple}\tiny{±0.047} & 0.77\color{darkpurple}\tiny{±0.036}
& 0.08\color{darkpurple}\tiny{±0.056} & 0.60\color{darkpurple}\tiny{±0.048}
& 0.05\color{darkpurple}\tiny{±0.052} & 0.55\color{darkpurple}\tiny{±0.045}
& \textbf{0.46}\color{darkpurple}\tiny{±0.038} & \textbf{0.81}\color{darkpurple}\tiny{±0.048} \\

\midrule
\rowcolor{TFRowGray}
\multicolumn{18}{c}{\textbf{TF+500}} \\
\midrule

\multirow{7}{*}{Specific}
& hESC
& 0.51\color{darkpurple}\tiny{±0.004} & 0.75\color{darkpurple}\tiny{±0.001}
& 0.55\color{darkpurple}\tiny{±0.002} & 0.78\color{darkpurple}\tiny{±0.005}
& 0.54\color{darkpurple}\tiny{±0.005} & 0.83\color{darkpurple}\tiny{±0.003}
& \underline{0.59}\color{darkpurple}\tiny{±0.001} & \underline{0.85}\color{darkpurple}\tiny{±0.004}
& 0.58\color{darkpurple}\tiny{±0.003} & 0.84\color{darkpurple}\tiny{±0.003}
& 0.23\color{darkpurple}\tiny{±0.005} & 0.52\color{darkpurple}\tiny{±0.002}
& 0.35\color{darkpurple}\tiny{±0.001} & 0.53\color{darkpurple}\tiny{±0.004}
& \textbf{0.61}\color{darkpurple}\tiny{±0.004} & \textbf{0.88}\color{darkpurple}\tiny{±0.001} \\

& hHEP
& 0.68\color{darkpurple}\tiny{±0.002} & 0.82\color{darkpurple}\tiny{±0.004}
& 0.81\color{darkpurple}\tiny{±0.005} & 0.79\color{darkpurple}\tiny{±0.001}
& 0.78\color{darkpurple}\tiny{±0.003} & 0.85\color{darkpurple}\tiny{±0.003}
& \underline{0.83}\color{darkpurple}\tiny{±0.003} & \underline{0.86}\color{darkpurple}\tiny{±0.003}
& 0.83\color{darkpurple}\tiny{±0.004} & 0.86\color{darkpurple}\tiny{±0.002}
& 0.48\color{darkpurple}\tiny{±0.001} & 0.53\color{darkpurple}\tiny{±0.005}
& 0.52\color{darkpurple}\tiny{±0.003} & 0.56\color{darkpurple}\tiny{±0.001}
& \textbf{0.84}\color{darkpurple}\tiny{±0.002} & \textbf{0.89}\color{darkpurple}\tiny{±0.004} \\

& mDC
& 0.11\color{darkpurple}\tiny{±0.003} & 0.65\color{darkpurple}\tiny{±0.001}
& 0.11\color{darkpurple}\tiny{±0.003} & 0.68\color{darkpurple}\tiny{±0.004}
& 0.15\color{darkpurple}\tiny{±0.005} & 0.72\color{darkpurple}\tiny{±0.002}
& \underline{0.19}\color{darkpurple}\tiny{±0.002} & \underline{0.73}\color{darkpurple}\tiny{±0.003}
& 0.12\color{darkpurple}\tiny{±0.004} & 0.70\color{darkpurple}\tiny{±0.003}
& 0.05\color{darkpurple}\tiny{±0.001} & 0.52\color{darkpurple}\tiny{±0.005}
& 0.08\color{darkpurple}\tiny{±0.003} & 0.49\color{darkpurple}\tiny{±0.002}
& \textbf{0.19}\color{darkpurple}\tiny{±0.004} & \textbf{0.76}\color{darkpurple}\tiny{±0.001} \\

& mESC
& 0.76\color{darkpurple}\tiny{±0.001} & 0.83\color{darkpurple}\tiny{±0.004}
& 0.82\color{darkpurple}\tiny{±0.005} & 0.86\color{darkpurple}\tiny{±0.002}
& 0.77\color{darkpurple}\tiny{±0.002} & \underline{0.89}\color{darkpurple}\tiny{±0.003}
& \underline{0.84}\color{darkpurple}\tiny{±0.004} & \underline{0.89}\color{darkpurple}\tiny{±0.001}
& 0.83\color{darkpurple}\tiny{±0.003} & 0.88\color{darkpurple}\tiny{±0.005}
& 0.78\color{darkpurple}\tiny{±0.003} & 0.53\color{darkpurple}\tiny{±0.002}
& 0.47\color{darkpurple}\tiny{±0.004} & 0.55\color{darkpurple}\tiny{±0.003}
& \textbf{0.85}\color{darkpurple}\tiny{±0.002} & \textbf{0.92}\color{darkpurple}\tiny{±0.004} \\

& mHSC-E
& 0.89\color{darkpurple}\tiny{±0.005} & 0.88\color{darkpurple}\tiny{±0.001}
& 0.94\color{darkpurple}\tiny{±0.002} & 0.90\color{darkpurple}\tiny{±0.004}
& 0.91\color{darkpurple}\tiny{±0.004} & 0.89\color{darkpurple}\tiny{±0.003}
& 0.93\color{darkpurple}\tiny{±0.001} & 0.89\color{darkpurple}\tiny{±0.003}
& \underline{0.94}\color{darkpurple}\tiny{±0.003} & \underline{0.91}\color{darkpurple}\tiny{±0.002}
& 0.89\color{darkpurple}\tiny{±0.005} & 0.54\color{darkpurple}\tiny{±0.001}
& 0.77\color{darkpurple}\tiny{±0.003} & 0.52\color{darkpurple}\tiny{±0.004}
& \textbf{0.94}\color{darkpurple}\tiny{±0.002} & \textbf{0.91}\color{darkpurple}\tiny{±0.003} \\

& mHSC-GM
& 0.89\color{darkpurple}\tiny{±0.003} & 0.88\color{darkpurple}\tiny{±0.001}
& 0.93\color{darkpurple}\tiny{±0.005} & 0.90\color{darkpurple}\tiny{±0.002}
& 0.90\color{darkpurple}\tiny{±0.001} & 0.89\color{darkpurple}\tiny{±0.004}
& 0.92\color{darkpurple}\tiny{±0.004} & 0.90\color{darkpurple}\tiny{±0.003}
& \textbf{0.94}\color{darkpurple}\tiny{±0.002} & \textbf{0.91}\color{darkpurple}\tiny{±0.005}
& 0.87\color{darkpurple}\tiny{±0.003} & 0.51\color{darkpurple}\tiny{±0.003}
& 0.78\color{darkpurple}\tiny{±0.004} & 0.54\color{darkpurple}\tiny{±0.001}
& \underline{0.93}\color{darkpurple}\tiny{±0.001} & \textbf{0.91}\color{darkpurple}\tiny{±0.002} \\

& mHSC-L
& 0.82\color{darkpurple}\tiny{±0.002} & 0.81\color{darkpurple}\tiny{±0.004}
& 0.86\color{darkpurple}\tiny{±0.005} & 0.86\color{darkpurple}\tiny{±0.001}
& 0.84\color{darkpurple}\tiny{±0.001} & 0.85\color{darkpurple}\tiny{±0.003}
& \underline{0.86}\color{darkpurple}\tiny{±0.004} & \underline{0.89}\color{darkpurple}\tiny{±0.003}
& \underline{0.86}\color{darkpurple}\tiny{±0.003} & 0.89\color{darkpurple}\tiny{±0.005}
& 0.58\color{darkpurple}\tiny{±0.003} & 0.61\color{darkpurple}\tiny{±0.002}
& 0.67\color{darkpurple}\tiny{±0.004} & 0.52\color{darkpurple}\tiny{±0.001}
& \textbf{0.87}\color{darkpurple}\tiny{±0.002} & \textbf{0.90}\color{darkpurple}\tiny{±0.004} \\

\hline

\multirow{7}{*}{Non-Specific}
& hESC
& \textbf{0.04}\color{darkpurple}\tiny{±0.009} & 0.67\color{darkpurple}\tiny{±0.008}
& \textbf{0.04}\color{darkpurple}\tiny{±0.010} & 0.67\color{darkpurple}\tiny{±0.009}
& \textbf{0.04}\color{darkpurple}\tiny{±0.008} & \underline{0.68}\color{darkpurple}\tiny{±0.007}
& \textbf{0.04}\color{darkpurple}\tiny{±0.011} & \underline{0.68}\color{darkpurple}\tiny{±0.009}
& \textbf{0.04}\color{darkpurple}\tiny{±0.006} & \underline{0.68}\color{darkpurple}\tiny{±0.008}
& \underline{0.02}\color{darkpurple}\tiny{±0.010} & 0.66\color{darkpurple}\tiny{±0.009}
& \underline{0.02}\color{darkpurple}\tiny{±0.009} & 0.66\color{darkpurple}\tiny{±0.008}
& \textbf{0.04}\color{darkpurple}\tiny{±0.004} & \textbf{0.69}\color{darkpurple}\tiny{±0.006} \\

& hHEP
& \textbf{0.05}\color{darkpurple}\tiny{±0.012} & 0.69\color{darkpurple}\tiny{±0.010}
& \textbf{0.05}\color{darkpurple}\tiny{±0.011} & \underline{0.70}\color{darkpurple}\tiny{±0.012}
& \textbf{0.05}\color{darkpurple}\tiny{±0.009} & \underline{0.71}\color{darkpurple}\tiny{±0.010}
& \textbf{0.05}\color{darkpurple}\tiny{±0.013} & \underline{0.70}\color{darkpurple}\tiny{±0.011}
& \textbf{0.05}\color{darkpurple}\tiny{±0.008} & 0.69\color{darkpurple}\tiny{±0.009}
& \underline{0.02}\color{darkpurple}\tiny{±0.011} & 0.61\color{darkpurple}\tiny{±0.012}
& \underline{0.02}\color{darkpurple}\tiny{±0.010} & 0.58\color{darkpurple}\tiny{±0.011}
& \textbf{0.05}\color{darkpurple}\tiny{±0.006} & \textbf{0.71}\color{darkpurple}\tiny{±0.009} \\

& mDC
& 0.15\color{darkpurple}\tiny{±0.041} & 0.64\color{darkpurple}\tiny{±0.011}
& 0.16\color{darkpurple}\tiny{±0.039} & 0.73\color{darkpurple}\tiny{±0.012}
& 0.16\color{darkpurple}\tiny{±0.043} & 0.71\color{darkpurple}\tiny{±0.013}
& \underline{0.19}\color{darkpurple}\tiny{±0.030} & \underline{0.75}\color{darkpurple}\tiny{±0.009}
& 0.14\color{darkpurple}\tiny{±0.035} & 0.72\color{darkpurple}\tiny{±0.010}
& 0.03\color{darkpurple}\tiny{±0.040} & 0.51\color{darkpurple}\tiny{±0.013}
& 0.03\color{darkpurple}\tiny{±0.038} & 0.55\color{darkpurple}\tiny{±0.012}
& \textbf{0.19}\color{darkpurple}\tiny{±0.037} & \textbf{0.78}\color{darkpurple}\tiny{±0.006} \\

& mESC
& 0.04\color{darkpurple}\tiny{±0.020} & 0.75\color{darkpurple}\tiny{±0.028}
& \textbf{0.06}\color{darkpurple}\tiny{±0.011} & \underline{0.76}\color{darkpurple}\tiny{±0.030}
& \textbf{0.08}\color{darkpurple}\tiny{±0.019} & \textbf{0.77}\color{darkpurple}\tiny{±0.027}
& 0.07\color{darkpurple}\tiny{±0.022} & 0.75\color{darkpurple}\tiny{±0.029}
& \textbf{0.08}\color{darkpurple}\tiny{±0.015} & \textbf{0.77}\color{darkpurple}\tiny{±0.024}
& 0.02\color{darkpurple}\tiny{±0.060} & 0.56\color{darkpurple}\tiny{±0.031}
& 0.02\color{darkpurple}\tiny{±0.019} & 0.58\color{darkpurple}\tiny{±0.030}
& \underline{0.07}\color{darkpurple}\tiny{±0.016} & \underline{0.76}\color{darkpurple}\tiny{±0.026} \\

& mHSC-E
& 0.14\color{darkpurple}\tiny{±0.013} & \underline{0.74}\color{darkpurple}\tiny{±0.018}
& 0.13\color{darkpurple}\tiny{±0.041} & 0.73\color{darkpurple}\tiny{±0.019}
& 0.14\color{darkpurple}\tiny{±0.045} & \underline{0.74}\color{darkpurple}\tiny{±0.020}
& \textbf{0.17}\color{darkpurple}\tiny{±0.032} & \textbf{0.75}\color{darkpurple}\tiny{±0.013}
& \underline{0.15}\color{darkpurple}\tiny{±0.036} & \underline{0.74}\color{darkpurple}\tiny{±0.016}
& 0.03\color{darkpurple}\tiny{±0.042} & 0.61\color{darkpurple}\tiny{±0.020}
& 0.03\color{darkpurple}\tiny{±0.040} & 0.55\color{darkpurple}\tiny{±0.019}
& \textbf{0.17}\color{darkpurple}\tiny{±0.038} & \textbf{0.75}\color{darkpurple}\tiny{±0.014} \\

& mHSC-GM
& 0.22\color{darkpurple}\tiny{±0.040} & 0.68\color{darkpurple}\tiny{±0.031}
& 0.17\color{darkpurple}\tiny{±0.016} & 0.72\color{darkpurple}\tiny{±0.030}
& \underline{0.23}\color{darkpurple}\tiny{±0.042} & 0.73\color{darkpurple}\tiny{±0.033}
& \textbf{0.27}\color{darkpurple}\tiny{±0.031} & \underline{0.75}\color{darkpurple}\tiny{±0.027}
& \underline{0.23}\color{darkpurple}\tiny{±0.036} & 0.73\color{darkpurple}\tiny{±0.029}
& 0.04\color{darkpurple}\tiny{±0.039} & 0.54\color{darkpurple}\tiny{±0.034}
& 0.04\color{darkpurple}\tiny{±0.037} & 0.54\color{darkpurple}\tiny{±0.033}
& \textbf{0.27}\color{darkpurple}\tiny{±0.034} & \textbf{0.77}\color{darkpurple}\tiny{±0.028} \\

& mHSC-L
& 0.11\color{darkpurple}\tiny{±0.055} & 0.59\color{darkpurple}\tiny{±0.035}
& 0.15\color{darkpurple}\tiny{±0.050} & 0.63\color{darkpurple}\tiny{±0.037}
& 0.15\color{darkpurple}\tiny{±0.058} & 0.59\color{darkpurple}\tiny{±0.040}
& \underline{0.17}\color{darkpurple}\tiny{±0.045} & \underline{0.65}\color{darkpurple}\tiny{±0.030}
& 0.15\color{darkpurple}\tiny{±0.048} & 0.64\color{darkpurple}\tiny{±0.034}
& 0.05\color{darkpurple}\tiny{±0.052} & 0.53\color{darkpurple}\tiny{±0.039}
& 0.05\color{darkpurple}\tiny{±0.050} & 0.52\color{darkpurple}\tiny{±0.038}
& \textbf{0.18}\color{darkpurple}\tiny{±0.049} & \textbf{0.68}\color{darkpurple}\tiny{±0.032} \\

\hline

\multirow{7}{*}{STRING}
& hESC
& 0.17\color{darkpurple}\tiny{±0.019} & 0.82\color{darkpurple}\tiny{±0.016}
& 0.22\color{darkpurple}\tiny{±0.021} & 0.83\color{darkpurple}\tiny{±0.018}
& 0.20\color{darkpurple}\tiny{±0.017} & 0.82\color{darkpurple}\tiny{±0.019}
& \underline{0.24}\color{darkpurple}\tiny{±0.014} & \underline{0.85}\color{darkpurple}\tiny{±0.011}
& \underline{0.24}\color{darkpurple}\tiny{±0.016} & 0.86\color{darkpurple}\tiny{±0.013}
& 0.04\color{darkpurple}\tiny{±0.020} & 0.65\color{darkpurple}\tiny{±0.021}
& 0.06\color{darkpurple}\tiny{±0.018} & 0.68\color{darkpurple}\tiny{±0.019}
& \textbf{0.26}\color{darkpurple}\tiny{±0.013} & \textbf{0.87}\color{darkpurple}\tiny{±0.008} \\

& hHEP
& \underline{0.24}\color{darkpurple}\tiny{±0.021} & \underline{0.86}\color{darkpurple}\tiny{±0.018}
& 0.18\color{darkpurple}\tiny{±0.023} & 0.84\color{darkpurple}\tiny{±0.020}
& \textbf{0.25}\color{darkpurple}\tiny{±0.019} & \textbf{0.87}\color{darkpurple}\tiny{±0.017}
& 0.19\color{darkpurple}\tiny{±0.024} & 0.83\color{darkpurple}\tiny{±0.016}
& 0.19\color{darkpurple}\tiny{±0.018} & 0.82\color{darkpurple}\tiny{±0.015}
& 0.03\color{darkpurple}\tiny{±0.022} & 0.69\color{darkpurple}\tiny{±0.023}
& 0.06\color{darkpurple}\tiny{±0.020} & 0.71\color{darkpurple}\tiny{±0.021}
& 0.20\color{darkpurple}\tiny{±0.069} & \underline{0.86}\color{darkpurple}\tiny{±0.017} \\

& mDC
& 0.26\color{darkpurple}\tiny{±0.025} & 0.75\color{darkpurple}\tiny{±0.020}
& 0.34\color{darkpurple}\tiny{±0.028} & 0.81\color{darkpurple}\tiny{±0.022}
& \underline{0.34}\color{darkpurple}\tiny{±0.026} & \underline{0.87}\color{darkpurple}\tiny{±0.023}
& \underline{0.34}\color{darkpurple}\tiny{±0.019} & 0.87\color{darkpurple}\tiny{±0.016}
& 0.33\color{darkpurple}\tiny{±0.024} & 0.86\color{darkpurple}\tiny{±0.018}
& 0.07\color{darkpurple}\tiny{±0.027} & 0.73\color{darkpurple}\tiny{±0.026}
& 0.06\color{darkpurple}\tiny{±0.025} & 0.65\color{darkpurple}\tiny{±0.024}
& \textbf{0.36}\color{darkpurple}\tiny{±0.022} & \textbf{0.88}\color{darkpurple}\tiny{±0.009} \\

& mESC
& 0.13\color{darkpurple}\tiny{±0.028} & 0.78\color{darkpurple}\tiny{±0.026}
& 0.22\color{darkpurple}\tiny{±0.030} & 0.84\color{darkpurple}\tiny{±0.028}
& \underline{0.23}\color{darkpurple}\tiny{±0.027} & 0.84\color{darkpurple}\tiny{±0.029}
& \underline{0.23}\color{darkpurple}\tiny{±0.021} & \underline{0.86}\color{darkpurple}\tiny{±0.018}
&  \underline{0.23}\color{darkpurple}\tiny{±0.025} & 0.85\color{darkpurple}\tiny{±0.022}
& 0.06\color{darkpurple}\tiny{±0.029} & 0.70\color{darkpurple}\tiny{±0.030}
& 0.06\color{darkpurple}\tiny{±0.027} & 0.76\color{darkpurple}\tiny{±0.028}
& \textbf{0.25}\color{darkpurple}\tiny{±0.023} & \textbf{0.88}\color{darkpurple}\tiny{±0.004} \\

& mHSC-E
& 0.23\color{darkpurple}\tiny{±0.038} & 0.82\color{darkpurple}\tiny{±0.028}
& 0.29\color{darkpurple}\tiny{±0.041} & 0.81\color{darkpurple}\tiny{±0.031}
& 0.26\color{darkpurple}\tiny{±0.036} & 0.83\color{darkpurple}\tiny{±0.030}
& \underline{0.31}\color{darkpurple}\tiny{±0.029} & \underline{0.85}\color{darkpurple}\tiny{±0.021}
& 0.30\color{darkpurple}\tiny{±0.033} & 0.82\color{darkpurple}\tiny{±0.025}
& 0.10\color{darkpurple}\tiny{±0.040} & 0.59\color{darkpurple}\tiny{±0.034}
& 0.06\color{darkpurple}\tiny{±0.037} & 0.62\color{darkpurple}\tiny{±0.032}
& \textbf{0.33}\color{darkpurple}\tiny{±0.042} & \textbf{0.87}\color{darkpurple}\tiny{±0.010} \\

& mHSC-GM
& 0.36\color{darkpurple}\tiny{±0.046} & 0.83\color{darkpurple}\tiny{±0.037}
& 0.42\color{darkpurple}\tiny{±0.049} & 0.84\color{darkpurple}\tiny{±0.039}
& \textbf{0.46}\color{darkpurple}\tiny{±0.044} & \textbf{0.88}\color{darkpurple}\tiny{±0.041}
& 0.43\color{darkpurple}\tiny{±0.036} & 0.85\color{darkpurple}\tiny{±0.031}
& 0.43\color{darkpurple}\tiny{±0.040} & 0.86\color{darkpurple}\tiny{±0.034}
& 0.06\color{darkpurple}\tiny{±0.047} & 0.71\color{darkpurple}\tiny{±0.042}
& 0.09\color{darkpurple}\tiny{±0.045} & 0.70\color{darkpurple}\tiny{±0.040}
& \underline{0.44}\color{darkpurple}\tiny{±0.063} & \underline{0.87}\color{darkpurple}\tiny{±0.009} \\

& mHSC-L
& 0.28\color{darkpurple}\tiny{±0.058} & 0.78\color{darkpurple}\tiny{±0.043}
& 0.41\color{darkpurple}\tiny{±0.061} & 0.81\color{darkpurple}\tiny{±0.045}
& \textbf{0.43}\color{darkpurple}\tiny{±0.066} & \underline{0.82}\color{darkpurple}\tiny{±0.048}
& 0.39\color{darkpurple}\tiny{±0.051} & 0.81\color{darkpurple}\tiny{±0.002}
& \underline{0.41}\color{darkpurple}\tiny{±0.055} & 0.81\color{darkpurple}\tiny{±0.007}
& 0.09\color{darkpurple}\tiny{±0.060} & 0.73\color{darkpurple}\tiny{±0.013}
& 0.07\color{darkpurple}\tiny{±0.057} & 0.60\color{darkpurple}\tiny{±0.001}
& \textbf{0.43}\color{darkpurple}\tiny{±0.103} & \textbf{0.85}\color{darkpurple}\tiny{±0.019} \\

\hline

\end{tabular}
\end{adjustbox}

\label{tab:main_grn_21_comparison}

\end{table*}
\vspace{-6pt}

\begin{table*}[!t]
\centering
\caption{Ablation results of BRIDGE across seven cell types.}
\label{tab:srta_ablation}

\small
\fontfamily{ptm}\selectfont
\renewcommand{\arraystretch}{1.1}
\setlength{\tabcolsep}{5pt}

\begingroup
\begin{adjustbox}{max width=0.9\textwidth}
\rowcolors{3}{black!3}{white}
\begin{tabular}{ccccccccccc}
\toprule
\rowcolor{HeadPurple}
\multicolumn{4}{c}{Components} & \multicolumn{7}{c}{Cell types} \\
\cmidrule(lr){1-4}\cmidrule(lr){5-11}
\rowcolor{HeadPurple}
BER & gate & cell graph & cell context & hESC & hHEP & mDC & mESC & mHSC-E & mHSC-GM & mHSC-L \\
\midrule
\xmark & \cmark & \cmark & \cmark & 0.58 & 0.79 & 0.12 & 0.81 & 0.92 & 0.92 & 0.82 \\
\cmark & \xmark & \cmark & \cmark & 0.52 & 0.74 & 0.11 & 0.73 & 0.87 & 0.86 & 0.79 \\
\xmark & \cmark & \xmark & \cmark & 0.51 & 0.72 & 0.10 & 0.69 & 0.84 & 0.81 & 0.74 \\
\cmark & \cmark & \xmark & \xmark & 0.53 & 0.68 & 0.07 & 0.66 & 0.81 & 0.79 & 0.64 \\
\midrule
\rowcolor{black!6}
\cmark & \cmark & \cmark & \cmark & \textbf{0.61} & \textbf{0.83} & \textbf{0.14} & \textbf{0.86} & \textbf{0.95} & \textbf{0.94} & \textbf{0.87} \\
\bottomrule
\end{tabular}
\end{adjustbox}
\endgroup
\end{table*}

\begin{figure*}[t]
\centering
\includegraphics[width=0.245\textwidth]{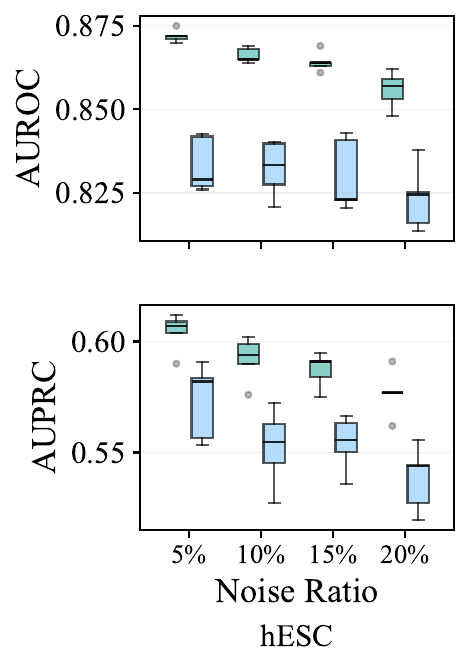}\hfill
\includegraphics[width=0.245\textwidth]{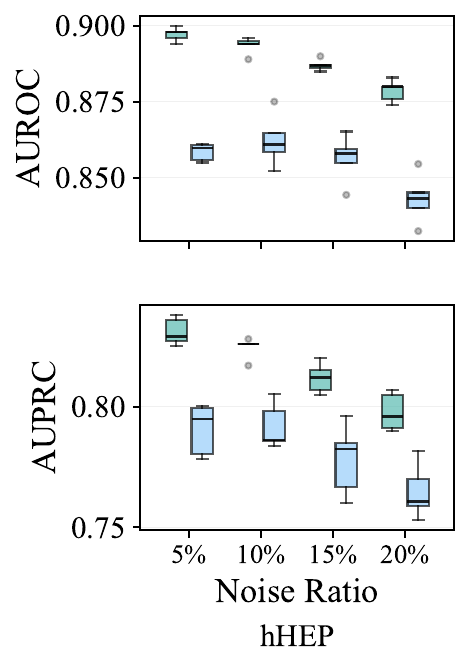}\hfill
\includegraphics[width=0.245\textwidth]{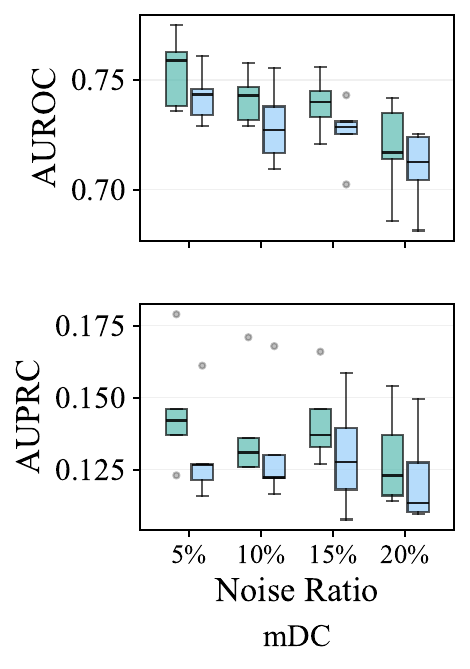}\hfill
\includegraphics[width=0.245\textwidth]{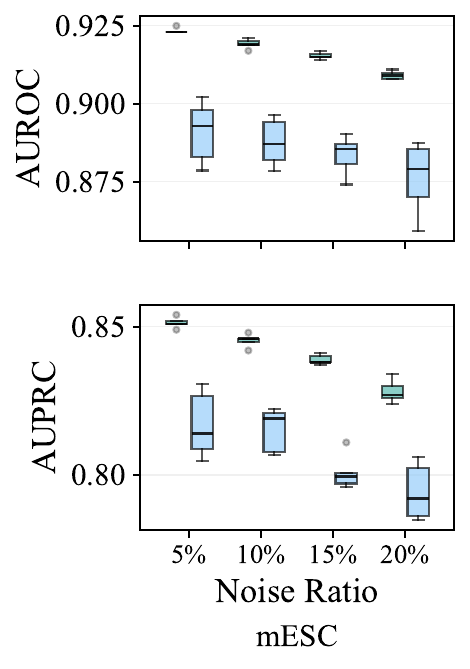}
\caption{BRIDGE GRN inference metrics (including AUROC above and AUPRC below) at different noise ratios for four cell types in a specific network. The green error bars represent BRIDGE, and the blue ones represent GCLink.}
\label{fig:noisy}
\end{figure*}


\subsection{GRN Inference Comparison on Benchmark}
We evaluate all methods on the GRN inference task under the same experimental setup), Specifically, We report AUPRC and AUROC, where AUPRC better reflects performance under severe class imbalance, while AUROC summarizes overall ranking quality. We compare BRIDGE with seven representative supervised GRN inference methods, including CNNC \citep{yuan2019deep}, GNE \citep{kc2019gne}, GENELink \citep{chen2022graph}, scTransNet \citep{kommu2024gene}, GMFGRN \citep{li2024gmfgrn}, HGATLink \citep{sun2025hgatlink}, and GCLink \citep{yu2025gclink}, covering diverse modeling paradigms. Detailed descriptions and implementation settings are provided in Appendix~\ref{baseline}.
In particular, Table \ref{tab:main_grn_21_comparison} shows the AUPRC and AUROC metrics of different methods for inferring gene regulatory relationships from scRNA-seq data. The results show that on a benchmark dataset covering three broad real-world regulatory network types and seven different cell types, BRIDGE outperforms robust baseline methods, achieving state-of-the-art AUROC and AUPRC indices across diverse settings in most cases.

In specific network data, BRIDGE markedly outperforms the second best method, GCLink \citep{yu2025gclink}, with an average AUPRC gain of 5\%. The improvement is primarily attributed to BRIDGE’s joint learning of gene and cell representations on heterogeneous graphs. To further examine this effect, we provide a aontrastive learning weight analysis in Appendix \ref{contrastive}, along with a performance analysis stratified by TF degree in Appendix \ref{degree}.
Compared with other heterogeneous graph methods such as HGATLink \citep{sun2025hgatlink} and GMFGRN \citep{li2024gmfgrn}, BRIDGE’s cross-type gating enhances generalization on Non-Specific networks, with larger gains under weak supervision and noisy regulatory priors. The advantage remains consistently evident on STRING and other Non-Specific settings despite reduced prior quality.Moreover, cell-conditioned decoding effectively attenuates unreliable cross-modal signals and maintains robust ranking performance across diverse regulatory scenarios in practice.
Furthermore, BRIDGE also yields consistently strong gains across various cell types, thus indicating robust performance under diverse regulatory contexts.
The larger gains in AUPRC than in AUROC suggest improved early precision, which better matches the sparse-positive nature of GRN discovery. 

\begin{figure*}[t]
\centering
\includegraphics[width=\textwidth]{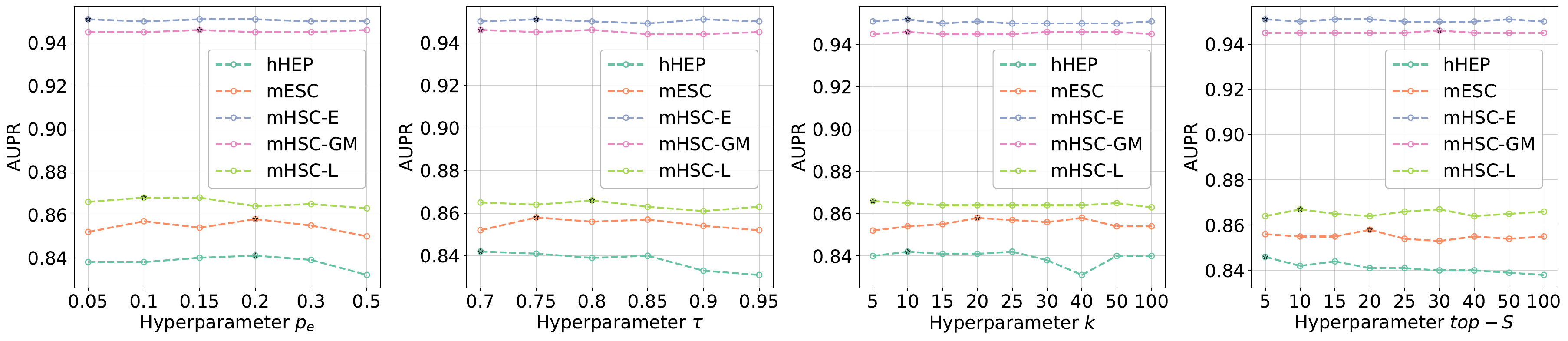}
\caption{AUPRC results of sensitivity testing of BRIDGE using a specific network dataset.}
\label{fig:ses}
\end{figure*}
\enlargethispage{\baselineskip}
\subsection{Ablation Studies}
We conduct ablation experiments on the BRIDGE framework to verify the contribution of each component to GRN inference. Based on the definitions in Section \ref{sec:method2}, we construct four control variants:
(i) \textbf{BRIDGE w/o BER}: Set $\tilde{A}^{gg}_{ij}=A^{gg}_{ij}$ in Eq. \eqref{eq:aug_refine}, using identical graph views without evidence refinement;
(ii) \textbf{BRIDGE w/o gate}: Set $\gamma^{g}_{i}=1,\gamma^{c}_{u}=1$ in Eq. \eqref{eq:update_gene}, yielding ungated cross-type message fusion;
(iii) \textbf{BRIDGE w/o cell graph and BER}: Set $\mathbf{A}^{cc}=\mathbf{I}$ in Eq.~\eqref{eq:acc_knn} and disable gene-cell neighborhood propagation in Eq.~\eqref{eq:agc_topk}. It retains the settings for removing the enhanced view;  
(iv) \textbf{BRIDGE w/o cell context and gate}: Based on the removal of the gating mechanism, it further drops $\mathbf{u}^{(v)}_i,\mathbf{u}^{(v)}_j$ in Eq. \eqref{eq:decoder_score}, decoding links from gene representations only. In this section's experiments, all settings are the same as the default settings in Section \ref{setup}, except for the component ablation treatments mentioned. All ablation experiments are performed on seven cell type datasets under the specifically regulated ChIP-seq network \citep{encode2020expanded}.

Table~\ref{tab:srta_ablation} shows that the BRIDGE model with all components achieves the highest performance on the cell-type-specific ChIP-seq dataset. Removing the evidence-guided augmentation (\textbf{BRIDGE w/o BER}) consistently reduces AUPRC, suggesting that biologically refined view construction is important for preserving informative regulatory signals under noisy supervision. Disabling cross-type gating (\textbf{BRIDGE w/o gate}) further degrades AUPRC, indicating that adaptive fusion is necessary to suppress noise introduced by gene-cell message exchange. When we simultaneously drop the cell graph and BER (\textbf{BRIDGE w/o cell graph and BER}), AUPRC decreases more sharply, highlighting the complementary roles of cell neighborhood propagation and evidence refinement. Finally, removing both cell context and gating (\textbf{BRIDGE w/o cell context and gate}) yields the lowest AUPRC, showing that cell-state conditioning in decoding is critical for accurate, context-aware regulation scoring. The results of these ablation experiments demonstrate that the components of BRIDGE work together in a complementary and coordinated manner to identify regulatory relationships from scRNA-seq data.

\subsection{Sensitivity Analysis to Noise}
\label{noisy}
Real-world GRN data often contains missing and spurious regulatory information \citep{tran2013counting, campos2019evolutionary}. To verify BRIDGE's robustness to noise in this context, we introduce random perturbations into the edges of the original GRN. We vary the perturbation rate to simulate different noise levels while keeping all other settings fixed. Experiments are conducted on specific regulatory networks and their corresponding seven cell types. Known edges are removed or spurious edges are added with fixed probabilities, enabling BRIDGE to infer gene regulatory relationships in an environment rich in structural noise. Figure ~\ref{fig:noisy} shows that although the gene regulatory relationship prediction performance of BRIDGE and GCLink \citep{yu2025gclink} decreases slightly with increasing noise, BRIDGE still maintains a higher accuracy. This trend indicates that BRIDGE is robust to both edge deletion and spurious edge injection. Noise sensitivity results for other cell types are in Appendix \ref{more_result}.

\enlargethispage{\baselineskip}
\subsection{Few-shot Studies}
\label{subsec:few-shot}
Due to the limited number of known gene regulatory interactions \citep{karamveer2024approaches}, only a small number of regulatory relationships are available for a specific cell type or cellular state. We adopt a cross-cell-type transfer learning strategy. Following the GCLink \citep{yu2025gclink} settings for transfer learning, we use mESC as the source cell type, and the other six cell types serve as target domains. For each target cell type dataset, we use 5\% of the data to fine tune the pretrained source model and use the remaining 95\% to evaluate AUROC and AUPRC. 
Results in Table~\ref{tab:few_shot_results} show consistent gains of BRIDGE over GCLink \citep{yu2025gclink} and GENELink \citep{chen2022graph} across all six target cell types on both AUPRC and AUROC. The margins are  clear on hESC and mHSC-GM, indicating stable cross-cell-type transfer with scarce supervision. Comparisons of Biological Evidence Refinement for view augmentation versus random edge removal, and Positive Sample Ratio Robustness Analysis, are reported in Appendix \ref{compare} and \ref{compare2}.

\subsection{Hyperparameter Analysis}
We perform sensitivity analyses of hyperparameters (Appendix~\ref{subsec:hyperparams}) on the ChIP-seq Specific network datasets \citep{encode2020expanded}. Specifically, we vary the evidence pruning ratio $p_e$ and quantile level $\rho$ in BER, the $k$-nearest neighbor size to construct $\mathbf{A}^{cc}$ for cell graph propagation in heterogeneous gated learning, and the neighborhood size $S$ for selecting similar cells in dual-space contrastive learning. Figure ~\ref{fig:ses} shows performance trends, highlighting how these choices affect robustness and accuracy across cell types.


\textbf{Pruning ratio $p_e$.} BRIDGE shows largely stable performance across different $p_e$ values, and the AUPRC only slightly decreases when the pruning ratio becomes large ($p_e=35\%$). This indicates that the biologically based edge removal enhancement method can effectively preserve high-quality structural signals for contrastive learning, even in the presence of substantial missing structural information.

\textbf{Quantile level $\rho$.} The selection of $\rho$ has little influence on BRIDGE, with AUPRC remaining consistently stable 
\enlargethispage{1\baselineskip}
across a wide range of quantile levels. The expression binarization step is stable under different global thresholds during coactivation estimation.


\textbf{$k$-nearest neighbor.} BRIDGE shows only minor fluctuations across different $k$ values, and the cell adjacency construction remains robust. The best performance appears at smaller neighborhood sizes (low $k$ values around 10), This indicates that expanding the neighborhood indiscriminately may introduce irrelevant or noisy cellular contexts. A compact local cell graph captures the most informative cell-state structure under noisy evidence. 

\begin{table}[t]
\centering
\small
\fontfamily{ptm}\selectfont
\renewcommand{\arraystretch}{1.1}
\setlength{\tabcolsep}{5pt}
\captionsetup{skip=5pt}
\caption{Few-shot Study Results: AUPR and AUC for Different Methods across Cell Types.}
\rowcolors{3}{black!3}{white}
\begin{tabular*}{0.95\columnwidth}{@{\hspace{2pt}}l@{\extracolsep{\fill}}cc@{\hspace{8pt}}cc@{\hspace{8pt}}cc@{\hspace{2pt}}}
\toprule
\rowcolor{HeadPurple}
\textbf{Cell Type} & \multicolumn{2}{c}{\textbf{BRIDGE}} & \multicolumn{2}{c}{\textbf{GCLink}} & \multicolumn{2}{c}{\textbf{GENELink}} \\
\cline{2-3}\cline{4-5}\cline{6-7}
\rowcolor{HeadPurple}
& \textbf{AUPR} & \textbf{AUC} & \textbf{AUPR} & \textbf{AUC} & \textbf{AUPR} & \textbf{AUC} \\
\midrule
hESC    & 0.44 & 0.80 & 0.39 & 0.79 & 0.38 & 0.77 \\
hHEP    & 0.65 & 0.80 & 0.64 & 0.77 & 0.64 & 0.76 \\
mDC     & 0.10 & 0.74 & 0.10 & 0.75 & 0.09 & 0.73 \\
mHSC-E  & 0.86 & 0.85 & 0.86 & 0.83 & 0.85 & 0.82 \\
mHSC-GM & 0.84 & 0.83 & 0.82 & 0.82 & 0.80 & 0.78 \\
mHSC-L  & 0.70 & 0.74 & 0.68 & 0.79 & 0.66 & 0.70 \\
\bottomrule
\end{tabular*}
\label{tab:few_shot_results}
\end{table}

\textbf{Neighborhood size $S$.} AUPRC remains nearly flat across a wide range of $S$ values for all cell types. Small to moderate neighborhoods achieve the strongest overall performance, with $S=5$ already capturing most of the benefits, while increasing to a larger neighborhood (e.g., $S=35$) yields only limited additional gains in representation consistency. These results highlight that BRIDGE can flexibly handle different neighborhood sizes $S$ during decoding.

\subsection{Case Study}
To further evaluate the biological relevance of the inferred regulatory relationships, we conducted a case study on the \textit{hESC} dataset under the \textit{Specific} network setting using the TFs+1000 configuration. Specifically, all predicted TF--target interactions were ranked according to their confidence scores, and the top 500 candidate pairs were selected for downstream validation. Interactions that appeared in the training and validation sets were excluded, resulting in 215 novel candidate regulatory relationships. Detailed experimental settings are provided in Appendix~\ref{setup}.

\paragraph{Validation using ChIPBase.}
Under the \textit{hg38} human genome setting, we compared the predicted interactions with the ChIPBase database for validation. The results show that 9 out of the top 10 predicted novel interactions are supported by ChIPBase (Table~\ref{tab:top10_pairs_hesc}), with most of these interactions associated with the transcription factor \textit{TFAP2A}, indicating high precision among top-ranked predictions.
\begin{figure*}[!t]
  \centering
  \includegraphics[width=\textwidth]{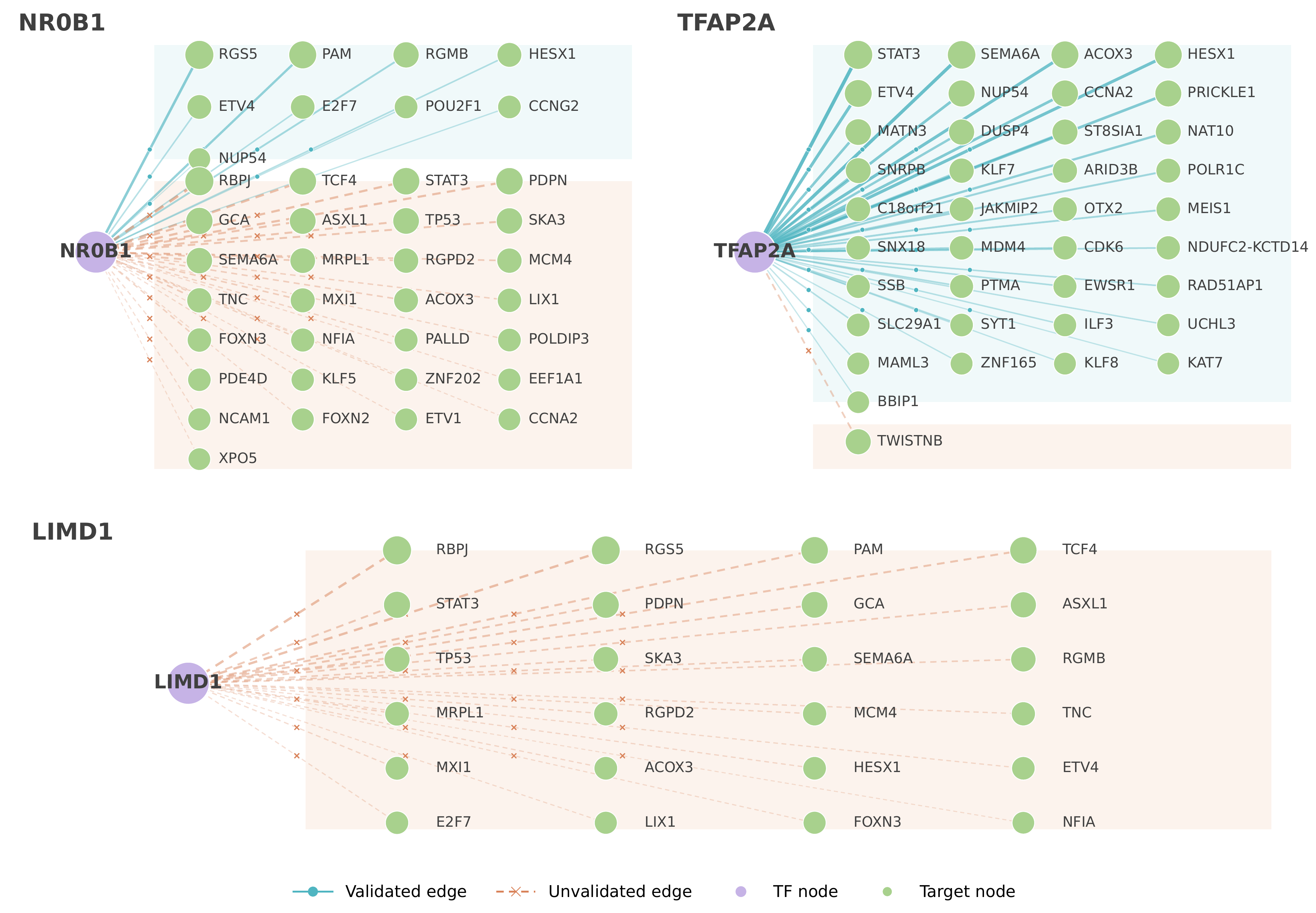}
  \caption{Validation results of the ChIPBase dataset dominated by TF genes.}
  \label{fig:case_study1}
\end{figure*}

As the number of predictions increases, the validation rate decreases but remains relatively high overall. Specifically, 46 out of the top 100 predicted interactions are supported by ChIPBase (Figure~\ref{fig:case_study1}). Further analysis reveals notable differences across transcription factors. \textit{TFAP2A} shows strong agreement with existing experimental evidence, with 37 out of 38 predicted targets being supported. In contrast, \textit{NR0B1} exhibits a moderate validation rate (9 out of 38), while none of the 24 predicted targets of \textit{LIMD1} are supported by current databases.

\begin{table}[t]
\centering
\small
\fontfamily{ptm}\selectfont
\renewcommand{\arraystretch}{1.1}
\setlength{\tabcolsep}{5pt}
\caption{Top-10 novel TF--target pairs predicted by BRIDGE on the hESC dataset.}
\rowcolors{2}{black!3}{white}
\begin{tabular*}{\columnwidth}{@{\hspace{2pt}}l@{\extracolsep{\fill}}l l@{\hspace{2pt}}}
\toprule
\rowcolor{HeadPurple}
\textbf{TF} & \textbf{Target gene} & \textbf{Reference} \\
\midrule
TFAP2A & STAT3    & ChIPBase \citep{huang2023chipbase} \\
TFAP2A & SEMA6A   & ChIPBase \citep{huang2023chipbase} \\
TFAP2A & ACOX3    & ChIPBase \citep{huang2023chipbase} \\
TFAP2A & HESX1    & ChIPBase \citep{huang2023chipbase} \\
TFAP2A & ETV4     & ChIPBase \citep{huang2023chipbase} \\
TFAP2A & NUP54    & ChIPBase \citep{huang2023chipbase} \\
TFAP2A & CCNA2    & ChIPBase \citep{huang2023chipbase} \\
TFAP2A & PRICKLE1 & ChIPBase \citep{huang2023chipbase} \\
NR0B1  & RBPJ     & -- \\
TFAP2A & MATN3    & ChIPBase \citep{huang2023chipbase} \\
\bottomrule
\end{tabular*}
\label{tab:top10_pairs_hesc}
\end{table}

\begin{figure*}[!t]
  \centering
  \includegraphics[width=\textwidth]{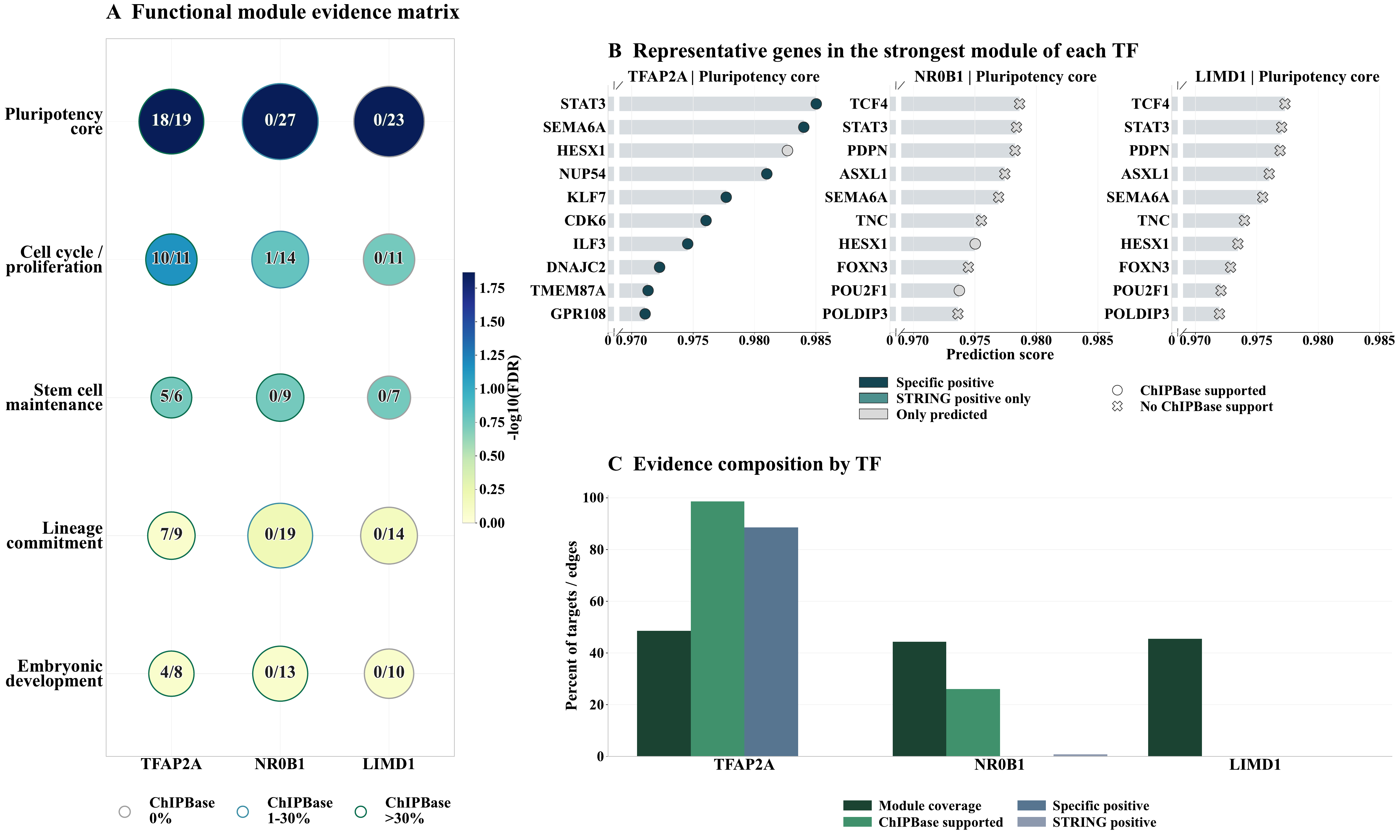}
  \caption{TF functional analysis diagrams: Figure A: Evidence of predicted target genes of each transcription factor at the functional module level. Figure B: Representative high-scoring target genes in the functional modules and their external support. Figure C: Module coverage and evidence composition of each transcription factor.}
  \label{fig:case_study2}
\end{figure*}

\paragraph{Functional evidence and biological interpretation.}
To further interpret these results, we performed functional analysis of the inferred regulatory relationships. Specifically, the candidate gene pairs were grouped by transcription factor, and three representative TFs (\textit{TFAP2A}, \textit{NR0B1}, and \textit{LIMD1}) were selected for in-depth analysis. Functional enrichment analysis was conducted using gene sets from the Molecular Signatures Database (MSigDB), focusing on key biological processes such as pluripotency, stem cell maintenance, cell proliferation, and lineage specification (results shown in Figure~\ref{fig:case_study2}).

The results indicate that all three TFs are significantly associated with pluripotency core and cell cycle/proliferation functions. Among them, the high validation rate of \textit{TFAP2A} is consistent with its well-established role in early lineage specification. Previous studies have shown that \textit{TFAP2A} plays a critical role in transcriptional reprogramming during the transition from pluripotency to differentiation, particularly in BMP-induced differentiation toward trophoblast and amnion lineages, where it represses pluripotency-associated factors such as \textit{OCT4} and \textit{NANOG}\citep{sekulovski2024temporally} \citep{castillo2023origin}. This functional role is consistent with the model predictions, suggesting that the inferred regulatory relationships capture biologically meaningful transcriptional programs.

For \textit{NR0B1}, as a known component of the pluripotency regulatory network, it participates in maintaining cell state through feedback interactions with core transcription factors such as \textit{OCT4} and \textit{ESRRB}\citep{uranishi2013dax1}.

In contrast, although \textit{LIMD1}-associated predictions lack direct support from current ChIPBase data, their target genes show enrichment in pluripotency and cell proliferation-related functions. This suggests that \textit{LIMD1} may play a potential role in the pluripotency regulatory network. Given that \textit{LIMD1} is known to act through signaling and post-transcriptional mechanisms that are less detectable by ChIP-based assays, we hypothesize that its regulatory effects may be mediated through non-canonical pathways, warranting further experimental validation.

\section{Discussion}

Our results demonstrate that incorporating joint heterogeneous modeling of genes and cells in single-cell data can effectively improve the performance and stability of gene regulatory network (GRN) inference. The biologically informed augmented views constructed in BRIDGE, together with contrastive learning in both gene and cell spaces, facilitate the capture of cross-type interactions and enhance the robustness of representation learning. In particular, the gating mechanism introduced in the heterogeneous encoding process adaptively regulates information propagation between genes and cells, mitigating noise propagation and enabling the model to accommodate diverse structural characteristics of real biological networks.

From the experimental results, BRIDGE achieves consistent performance improvements and strong generalization ability across multiple datasets. Meanwhile, the model also exhibits considerable biological interpretability. Further analysis shows that BRIDGE maintains predictive capability in low-data regimes, where limited target samples are available, by leveraging transferable structural and cross-domain information. In addition, the model demonstrates strong robustness to noisy or incomplete experimental data, as its performance does not degrade significantly under such conditions. Building upon these properties, the model is not only capable of recovering known regulatory relationships, but also demonstrates the ability to discover novel knowledge: after excluding edges observed in the training and validation sets, it can still identify biologically meaningful TF--target interactions and functional modules, suggesting that it captures underlying regulatory structures to a certain extent.

Despite these advantages, it is important to note that the inferred regulatory relationships are fundamentally statistical associations rather than direct evidence of causal regulation. Therefore, these predictions still require validation through external databases, experimental evidence, or biological priors. Future work will focus on integrating \textit{causal representation learning} and \textit{structural causal models (SCM)} into graph neural network frameworks to enhance the interpretability and causal reasoning capability of GRN inference. In addition, hyperparameter sensitivity analysis indicates that the model maintains stable performance within a reasonable range of key parameters (e.g., neighborhood size, contrastive loss weights, and edge perturbation ratios), suggesting a degree of robustness to parameter selection. Nevertheless, developing more adaptive parameter learning strategies remains an important direction for future research. Overall, how to balance model expressiveness with appropriate applicability, and how to effectively integrate computational predictions with experimental validation and domain knowledge, remains a key challenge for advancing GRN inference.

\section{Conflicts of interest}
The authors declare that they have no competing interests.

\section{Funding}
This work is supported in part by the National Natural Science Foundation of China under Grant 62306014 and 12501344, Postdoctoral Fellowship Program (Grade A) of CPSF under Grant BX20250376 and BX20240239, China Postdoctoral Science Foundation under Grant 2024M762201, Sichuan Science and Technology Program under Grant 2025ZNSFSC1506 and 2025ZNSFSC0808.

\section{Data availability}
The datasets used in this study follow the BEELINE benchmark setting and are derived from publicly available scRNA-seq datasets. The source code, processed datasets, and data preprocessing scripts supporting this study are available at \url{https://github.com/ShanwenTan/BRIDGE}.

\section{Author contributions statement}
Ziyang Dong implemented the model, conducted the experiments, analyzed the results, and drafted the manuscript. Shanwen Tan contributed to model design, implementation, experiments, and result analysis. Hengchuang Yin contributed to manuscript writing and organization. Yifan Wang contributed to result analysis and visualization. Siyu Yi, Jiancheng Lv, and Wei Ju supervised the study, contributed to study design and result interpretation, and revised the manuscript. All authors read and approved the final manuscript.
\section{Acknowledgments}
The authors thank the anonymous reviewers for their valuable suggestions.


\nocite{*}
\bibliographystyle{plainnat}
\bibliography{example_paper}
\clearpage
\onecolumn
\begin{appendices}
\raggedbottom 
\setlength{\parindent}{0pt}
\setlength{\parskip}{0pt}
\section{Supplementary Details of Materials and Methods}
\label{sec:appendix_details}
\subsection{Algorithm}
\label{app:alg}
\FloatBarrier
To provide a clearer procedural description of BRIDGE, we summarize the complete training workflow in Algorithm~\ref{alg:bridge}. The algorithm integrates heterogeneous graph construction, BER-based view augmentation, gated representation learning, dual-space contrastive regularization, and TF--target link prediction into a unified optimization process.
\begin{algorithm}[H]
\caption{The BRIDGE algorithm.}
\label{alg:bridge}
\begin{algorithmic}[1]
\Require Expression matrix $X$, gene adjacency $A^{gg}$, training pairs $\mathcal{D}$, hyperparameters
\Ensure Predicted regulatory scores $\hat{A}$

\State Construct cell features $X^c$ from $X$.
\State Build cell graph $A^{cc}$ using $k$-NN (Eq.~\eqref{eq:acc_knn}).
\State Construct gene--cell bipartite graph $A^{gc}$ (Eq.~\eqref{eq:agc_topk}).
\State Compute co-activation scores and obtain refined gene graph $\tilde{A}^{gg}$ (Eq.~\eqref{eq:aug_refine}).
\State Form two graph views $\mathcal{G}$ and $\tilde{\mathcal{G}}$.

\For{$t = 1$ to $T$}
    \State Encode $\mathcal{G}$ and $\tilde{\mathcal{G}}$ via heterogeneous encoder (Eq.~\eqref{eq:gat_message}).
    \State Perform dynamic gated aggregation (Eqs.~\eqref{eq:gating_coef}--\eqref{eq:update_gene}).
    \State Update gene--cell relations via dynamic top-$k$ reconstruction (Appendix~\ref{app:dyn}).
    \State Obtain gene embeddings $H_g^{(t)}$ and cell embeddings $H_c^{(t)}$.
    \State Compute gene-space contrastive loss $\mathcal{L}_{\mathrm{con}}^g$ (Eq.~\eqref{eq:con_loss}).
    \State Compute cell-space contrastive loss $\mathcal{L}_{\mathrm{con}}^c$ (Eq.~\eqref{eq:con_loss}).
    \State Compute link prediction scores $p_{tr}$ using decoder (Eq.~\eqref{eq:decoder_score}).

    \If{training}
        \State Compute link prediction loss $\mathcal{L}_{\mathrm{link}}$ (Eq.~\eqref{eq:link_loss}).
        \State $\mathcal{L} \leftarrow \lambda_{\mathrm{link}} \mathcal{L}_{\mathrm{link}} + \lambda_g \mathcal{L}_{\mathrm{con}}^{g} + \lambda_c \mathcal{L}_{\mathrm{con}}^{c}$.
        \State Update model parameters via back-propagation.
    \EndIf
\EndFor

\State \Return $\hat{A}$.
\end{algorithmic}
\end{algorithm}
\FloatBarrier
\subsection{Heterogeneous Graph Attention.}
\label{app:hgat}
To model relation-specific interactions in the heterogeneous graph, we adopt a relation-aware graph attention mechanism. For each relation type $\phi \in \{gg, gc, cg, cc\}$, the aggregation function $\mathrm{GAT}_\phi$ computes messages by attending over the neighbors of node $v$ under relation $\phi$.

Specifically, given node representations $\mathbf{h}_v$ and its neighbor $\mathbf{h}_u$, we first project them into a shared latent space:
\[
\mathbf{h}_v' = \mathbf{W}_\phi \mathbf{h}_v, \quad
\mathbf{h}_u' = \mathbf{W}_\phi \mathbf{h}_u.
\]
where $\mathbf{W}_\phi$ is a relation-specific transformation matrix.

The attention coefficient between node $v$ and its neighbor $u$ is then computed as:
\[
e_{vu}^{\phi}
=
\mathrm{LeakyReLU}
\left(
\mathbf{a}_\phi^\top
\left[
\mathbf{h}_v' \,\Vert\, \mathbf{h}_u'
\right]
\right).
\]
where $\mathbf{a}_\phi$ is a learnable attention vector. The normalized attention weights are obtained via softmax:
\[
\alpha_{vu}^{\phi}
=
\frac{
\exp(e_{vu}^{\phi})
}{
\sum_{k \in \mathcal{N}_\phi(v)} \exp(e_{vk}^{\phi})
}.
\]

Finally, the relation-specific message is aggregated as a weighted sum of neighbor representations:
\[
\mathbf{m}_v^\phi
=
\sum_{u \in \mathcal{N}_\phi(v)}
\alpha_{vu}^{\phi} \mathbf{h}_u'.
\]

This formulation allows the model to adaptively weigh different neighbors under each relation type, enabling fine-grained control over heterogeneous information propagation.
\subsection{Heterogeneous Message Passing with Dynamic Reconstruction}
\label{app:dyn}
Given initial gene features $\mathbf{X}^g$ and cell features $\mathbf{X}^c$, we first construct the initial heterogeneous graph with adjacency matrices $A^{gg}$, $A^{cc}$, and $A^{gc}$.

\textbf{(1) Initial gene--cell connections.}
The initial gene--cell adjacency $A^{gc,(0)}$ is constructed based on expression-derived relationships.

\textbf{(2) First-layer message passing.}
We perform heterogeneous message passing over the initial graph:
\begin{align*}
\mathbf{h}_i^{g,(1)} &= \sigma\!\left(\mathbf{h}_i^{g,(0)} + \mathbf{m}_i^{gg,(0)} + \gamma_i^{g,(0)} \mathbf{m}_i^{gc,(0)}\right).\\
\mathbf{h}_u^{c,(1)} &= \sigma\!\left(\mathbf{h}_u^{c,(0)} + \mathbf{m}_u^{cc,(0)} + \gamma_u^{c,(0)} \mathbf{m}_u^{cg,(0)}\right).
\end{align*}.
where $\mathbf{m}^{\phi,(0)}$ denotes relation-specific messages computed via $\mathrm{GAT}_\phi$ over the initial graph.

\textbf{(3) Dynamic graph reconstruction}
\begingroup
\setlength{\abovedisplayskip}{6pt}
\setlength{\belowdisplayskip}{6pt}
\setlength{\abovedisplayshortskip}{4pt}
\setlength{\belowdisplayshortskip}{4pt}

Based on the updated embeddings $\mathbf{h}_i^{g,(1)}$ and $\mathbf{h}_u^{c,(1)}$, we reconstruct gene--cell connections. We first compute cosine similarity:
\[
S_{iu}^{gc} =
\frac{(\mathbf{h}_i^{g,(1)})^\top \mathbf{h}_u^{c,(1)}}{\|\mathbf{h}_i^{g,(1)}\|_2 \, \|\mathbf{h}_u^{c,(1)}\|_2}.
\]

For each gene node $i$, we select the top-$k$ most similar cells:
\[
\mathcal{K}_i = \{u \mid S_{iu}^{gc} \text{ ranks among top-}k \text{ in } S_{i,:}^{gc}\}.
\]

and update the adjacency:
\[
A_{iu}^{gc,(1)} = \mathbb{I}(u \in \mathcal{K}_i).
\]

\endgroup

\textbf{(4) Second-layer message passing.}
Using the reconstructed adjacency $A^{gc,(1)}$, we perform a second round of heterogeneous message passing:

\begingroup
\setlength{\abovedisplayskip}{6pt}
\setlength{\belowdisplayskip}{6pt}
\setlength{\abovedisplayshortskip}{4pt}
\setlength{\belowdisplayshortskip}{4pt}
\setlength{\jot}{2pt}
\begin{align*}
\mathbf{h}_i^{g,(2)} &= \sigma\!\left(\mathbf{h}_i^{g,(1)} + \mathbf{m}_i^{gg,(1)} + \gamma_i^{g,(1)} \mathbf{m}_i^{gc,(1)}\right),\\
\mathbf{h}_u^{c,(2)} &= \sigma\!\left(\mathbf{h}_u^{c,(1)} + \mathbf{m}_u^{cc,(1)} + \gamma_u^{c,(1)} \mathbf{m}_u^{cg,(1)}\right).
\end{align*}.
\endgroup

where cross-type messages $\mathbf{m}^{gc,(1)}$ and $\mathbf{m}^{cg,(1)}$ are computed based on the reconstructed graph.

The final node representations $\mathbf{h}^{(2)}$ are then used for downstream prediction and contrastive learning.
\subsection{Dataset Statistics}
\label{app:datasetstats}
We summarize the statistical properties of all datasets used in our experiments in Table~\ref{tab:dataset_statistics}. The datasets are constructed from three types of regulatory networks, including Specific, Non-Specific, and STRING networks, following the preprocessing protocol of prior works. For each network type and cell type, we extract transcription factors (TFs) and select the top 500 or 1000 highly variable genes (HVGs) based on expression variability.

\begin{table}[H]
\centering
\caption{Statistical overview of single-cell transcriptomes and regulatory networks using TFs and the 500 (1000) most variable genes.}
\label{tab:dataset_statistics}

\fontsize{9}{10}\selectfont
\fontfamily{ptm}\selectfont
\setlength{\tabcolsep}{4pt}

\begingroup
\renewcommand{\arraystretch}{1.02}
\setlength{\extrarowheight}{0pt}
\setlength{\aboverulesep}{0.3pt}
\setlength{\belowrulesep}{0.3pt}
\setlength{\doublerulesep}{0.15pt}
\begin{adjustbox}{max width=\textwidth}
\begin{tabular}{@{} l l r r r >{\raggedright\arraybackslash}p{4.4cm} r @{}}
\toprule
\cellcolor{HeadPurple}{Dataset} & \cellcolor{HeadPurple}{Cell types} & \cellcolor{HeadPurple}{Cells} & \cellcolor{HeadPurple}{TFs} & \cellcolor{HeadPurple}{Genes} & \cellcolor{HeadPurple}{Pos/Neg} & \cellcolor{HeadPurple}{Density} \\
\midrule
\multirow{7}{*}{Specific}
 & hESC     & 758  & 34(34)  & 815(1260)  & 4545/26361 (7084/40822)      & 0.172(0.174) \\
 & hHEP     & 425  & 30(31)  & 874(1331)  & 9939/18471 (15558/29299)     & 0.538(0.531) \\
 & mDC      & 383  & 20(21)  & 443(684)   & 756/15644 (1193/26527)       & 0.048(0.045) \\
 & mESC     & 421  & 88(89)  & 977(1385)  & 29613/68859 (42795/101296)   & 0.430(0.422) \\
 & mHSC-E   & 1071 & 29(33)  & 691(1177)  & 11557/8830 (21975/17724)     & 1.309(1.240) \\
 & mHSC-GM  & 889  & 22(23)  & 618(1089)  & 7364/6518 (14135/11878)      & 1.130(1.190) \\
 & mHSC-L   & 847  & 16(16)  & 525(640)   & 4398/4546 (5180/5876)        & 0.967(0.882) \\
\midrule
\multirow{7}{*}{Non-Specific}
 & hESC     & 758  & 283(292) & 753(1138) & 3441/253806 (4617/406811)    & 0.014(0.011) \\
 & hHEP     & 425  & 322(332) & 825(1217) & 4129/300805 (5351/475053)    & 0.014(0.011) \\
 & mDC      & 383  & 250(254) & 634(969)  & 3067/201933 (3918/331362)    & 0.015(0.012) \\
 & mESC     & 421  & 516(522) & 890(1214) & 6893/570511 (8030/837088)    & 0.012(0.010) \\
 & mHSC-E   & 1071 & 144(147) & 442(674)  & 1425/99807 (1960/174881)     & 0.014(0.011) \\
 & mHSC-GM  & 889  & 82(88)   & 297(526)  & 743/50999 (1358/98170)       & 0.015(0.014) \\
 & mHSC-L   & 847  & 35(37)   & 164(192)  & 279/19286 (317/25250)        & 0.014(0.013) \\
\midrule
\multirow{7}{*}{STRING}
 & hESC     & 758  & 343(351) & 511(695)  & 4257/307530 (5149/489410)    & 0.014(0.011) \\
 & hHEP     & 425  & 409(414) & 646(874)  & 7523/379800 (9003/590055)    & 0.020(0.015) \\
 & mDC      & 383  & 264(273) & 479(664)  & 4815/211665 (5898/354462)    & 0.023(0.017) \\
 & mESC     & 421  & 495(499) & 638(785)  & 7762/546143 (8479/799402)    & 0.014(0.011) \\
 & mHSC-E   & 1071 & 156(161) & 291(413)  & 1371/108297 (1826/191857)    & 0.013(0.010) \\
 & mHSC-GM  & 889  & 92(100)  & 201(344)  & 748/57304 (1311/111789)      & 0.013(0.012) \\
 & mHSC-L   & 847  & 39(40)   & 70(81)    & 137/21664 (154/27486)        & 0.006(0.006) \\
\bottomrule
\end{tabular}
\end{adjustbox}
\endgroup
\end{table}

The number of cells is determined by the corresponding single-cell RNA-seq dataset, while the numbers of TFs and genes are obtained after filtering and HVG selection. Positive samples correspond to known TF--target regulatory interactions, and negative samples are generated from non-interacting pairs. We report the ratio of positive to negative samples as well as the resulting network density, defined as the proportion of observed regulatory edges among all possible TF--target pairs. This table provides an overview of dataset scale, sparsity, and heterogeneity across different experimental settings.

\FloatBarrier
\section{Detailed Experimental Setup}
\label{setup}

\subsection{Hyperparameter Settings}
\label{subsec:hyperparams}
\paragraph{Training protocol.}
During training, model parameters are updated using a mini-batch strategy with 20 training epochs and a batch size of 256.
The Adam optimizer is employed with an initial learning rate of $3 \times 10^{-3}$, together with a StepLR learning rate scheduler that decayed the learning rate by a factor of 0.99 after each epoch.
After each training epoch, the model is evaluated on the validation set using AUC and AUPR, with AUPR adopted as the primary criterion for model selection.
The model parameters corresponding to the best validation AUPR are saved.
In addition, to enable the model to better fit the observed regulatory data, a 20-epoch pretraining stage is introduced, during which only the link prediction objective is optimized and contrastive learning is disabled.

\paragraph{Shared model configuration.}
The shared model configuration across all three types of regulatory networks is as follows.
The encoder consists of two heterogeneous graph attention (Heterogeneous GAT) layers with multi-head attention, where the number of attention heads is set to $(3,3)$.
The hidden layer dimensions are configured as $128 \rightarrow 64 \rightarrow 64$, and the latent embedding dimension is set to 32.
LeakyReLU is used as the activation function with a negative slope of $\alpha = 0.2$, and the outputs of multiple attention heads are aggregated by concatenation.
Relationships between genes and cells are modeled via Top-$k$ aggregation over cells with $k=20$.
An expression-aware graph augmentation strategy is employed, with an edge drop rate of $p_e = 0.2$ and a high-expression quantile threshold of $\rho = 0.8$. During decoding, the neighborhood size is set to $S = 10$.
During training, the overall objective jointly optimizes the link prediction loss and the gene- and cell-level contrastive losses, weighted by $\lambda_{link}=0.8$, $\lambda_g=0.1$, and $\lambda_c=0.1$, respectively.

\paragraph{Network-specific parameter settings.}
For \textbf{Specific} networks, regulatory interactions are cell type-specific and exhibit strong consistency between gene expression patterns and true regulatory edges.
Accordingly, a relatively conservative structural modeling strategy is adopted, where the cell graph is constructed using a $k$-nearest neighbor ($k$NN) graph with $k=20$ to avoid excessive smoothing and the introduction of redundant cell information.

For \textbf{Non-Specific} networks, regulatory interactions are not restricted to individual cell types, resulting in pronounced expression heterogeneity.
To capture more stable cell similarity structures, the neighborhood size of the $k$NN cell graph is increased to $k=40$, together with the expression-aware edge removing mechanism to suppress cross-cell noise in regulatory edges.

For \textbf{STRING} networks, which primarily encode functional associations rather than direct transcriptional regulation, the networks are denser and exhibit higher noise levels.
Therefore, the same parameter configuration as that of the Non-Specific networks is adopted ($k=40$), enabling enhanced cell neighborhood modeling and expression-aware graph augmentation to improve robustness under high-noise conditions.

\subsection{Baseline Method}
\label{baseline}

\textbf{CNNC} \citep{yuan2019deep} transforms the joint expression of a gene pair across cells into an image-like 2D histogram and applies a convolutional neural network to classify potential regulatory relationships.

\textbf{GNE} \citep{kc2019gne} learns gene embeddings by jointly modeling network topology and gene expression attributes, and uses an MLP to aggregate these signals for interaction prediction in the embedding space. 

\textbf{GENELink} \citep{chen2022graph} formulates GRN inference as TF--target link prediction and uses a graph attention network to learn gene representations from observed TF--gene links and expression-derived features. 

\textbf{scTransNet} \citep{kommu2024gene} integrates contextual gene representations from a pre-trained single-cell transformer with a GRN graph encoder (GNN/GAT) and attentive pooling to perform supervised regulatory link prediction. 

\textbf{GMFGRN} \citep{li2024gmfgrn} performs GNN-based matrix factorization to learn gene embeddings and predicts TF--gene regulations using the inferred low-rank interaction structure. 

\textbf{HGATLink} \citep{sun2025hgatlink} combines heterogeneous graph attention with a simplified transformer (with matrix-decomposition-based embedding) to capture heterogeneous dependencies and long-range gene interactions for regulation prediction. 

\textbf{GCLink} \citep{yu2025gclink} is a graph contrastive link prediction framework that constructs augmented graph views via edge perturbations, learns gene representations with GAT, and optimizes a contrastive objective to improve generalization under limited known regulations.


\section{Additional Experiments}
\label{additional}
\subsection{Contrastive Learning Weight Analysis}
\label{contrastive}
This subsection studies the weight selection of the contrastive objective in BRIDGE. We tune the gene-space contrastive weight $\lambda_g$ and the cell-space contrastive weight $\lambda_c$. The experiments are conducted on seven different cell type datasets under a specific regulatory network. AUPRC and AUROC are the primary metrics. We plot contour heatmaps of both metrics over $(\lambda_c,\lambda_g)$. These plots quantify sensitivity, identify feasible ranges, and support the default hyperparameter choice.

\begin{figure*}[t]
\centering
\includegraphics[width=0.245\textwidth]{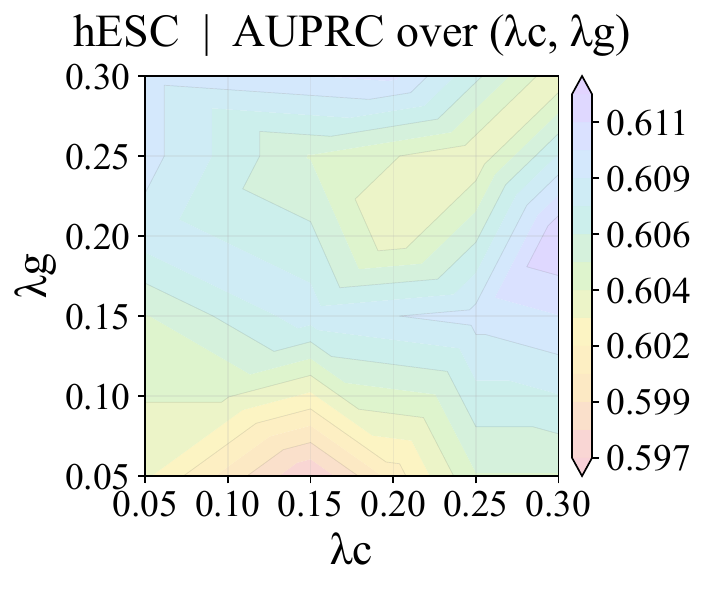}
\includegraphics[width=0.245\textwidth]{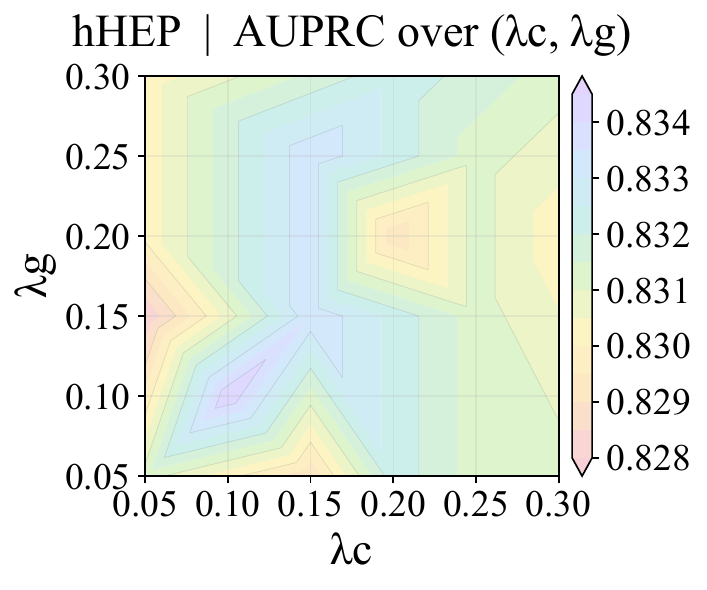}
\includegraphics[width=0.245\textwidth]{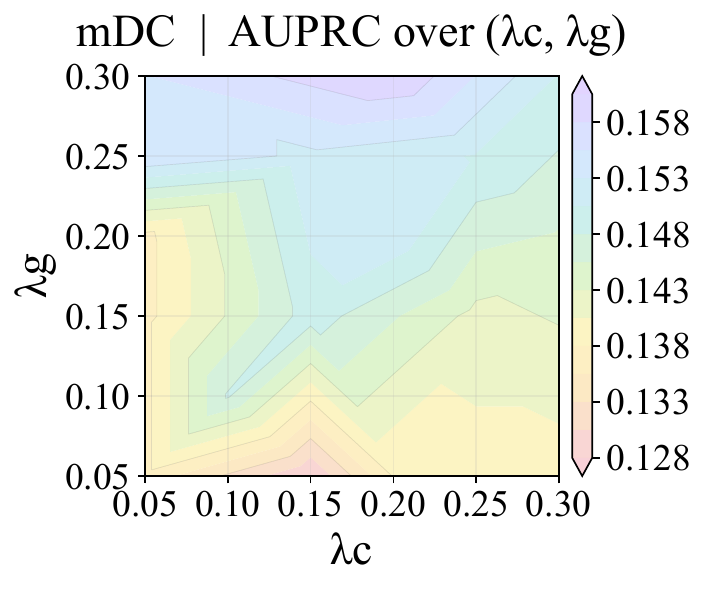}
\includegraphics[width=0.245\textwidth]{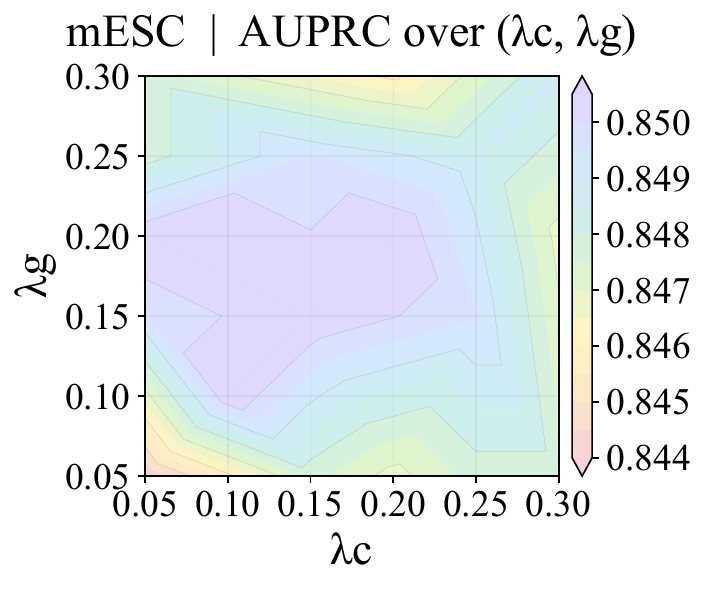}
\includegraphics[width=0.245\textwidth]{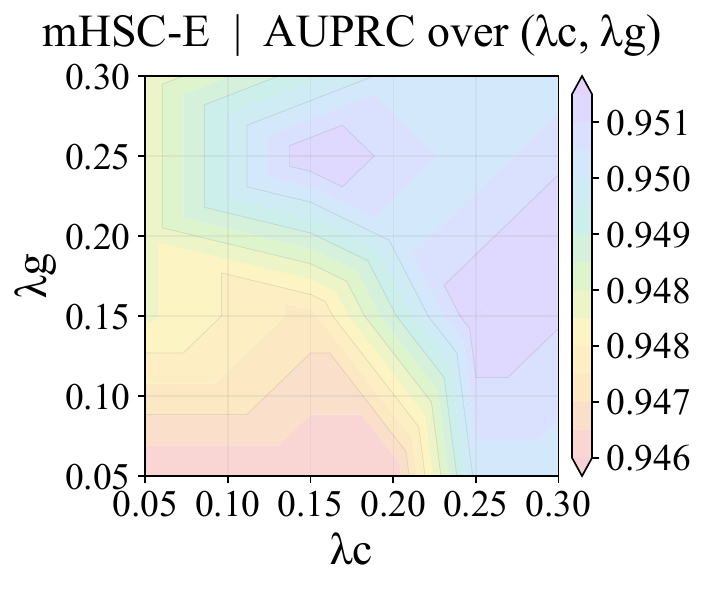}
\includegraphics[width=0.245\textwidth]{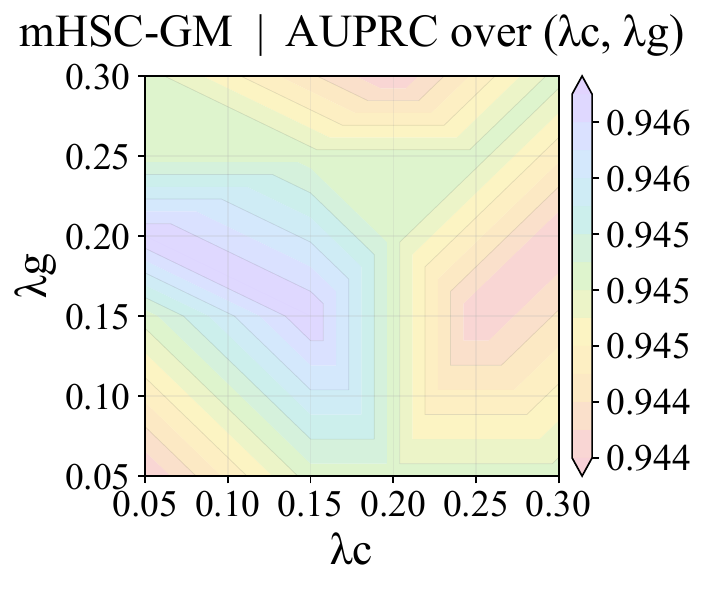}
\includegraphics[width=0.245\textwidth]{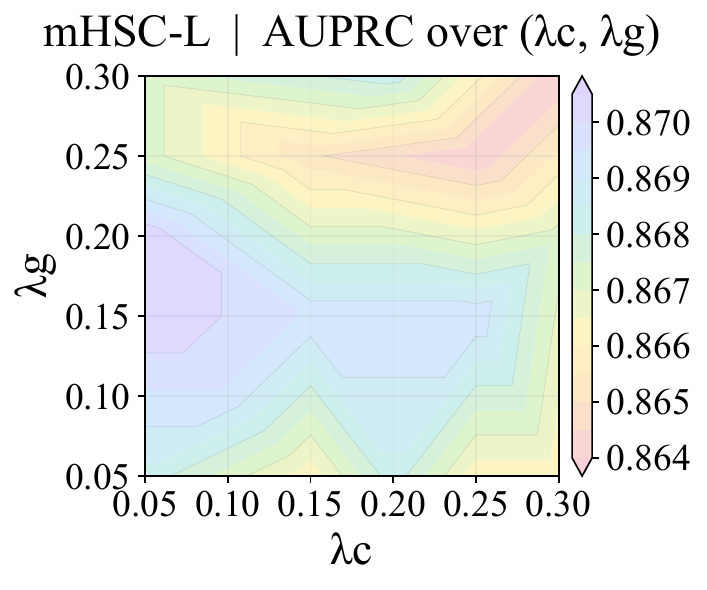}
\includegraphics[width=0.245\textwidth]{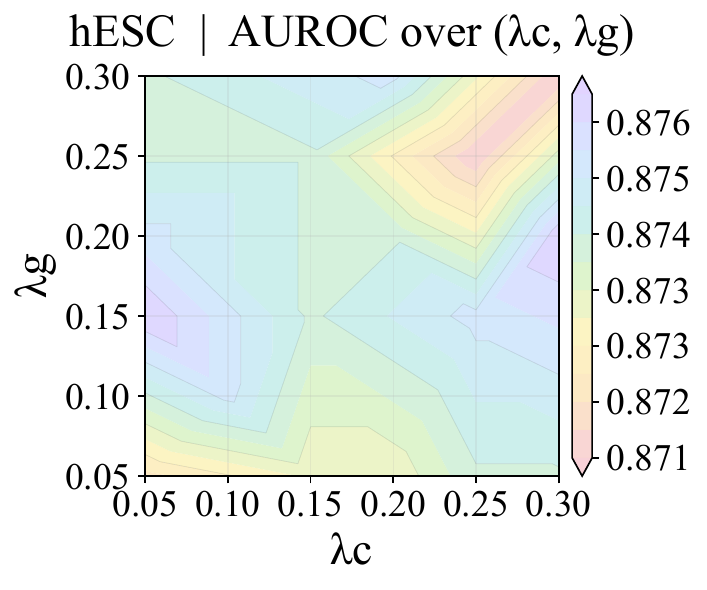}
\includegraphics[width=0.245\textwidth]{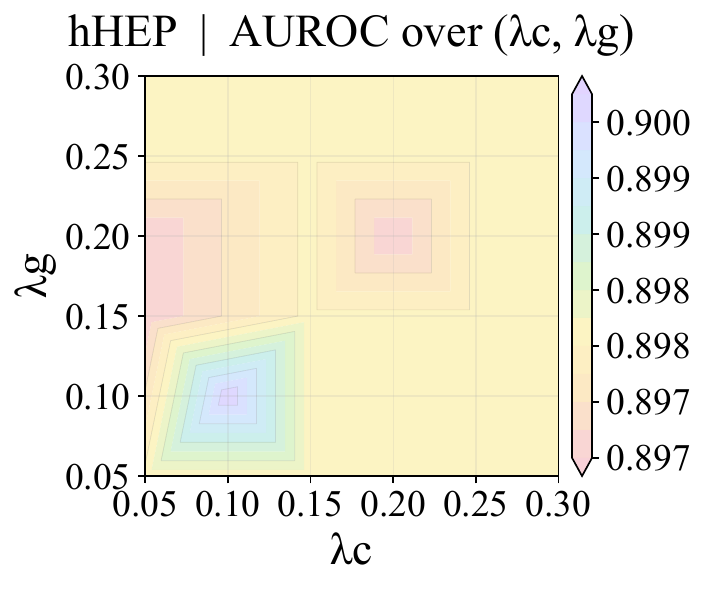}
\includegraphics[width=0.245\textwidth]{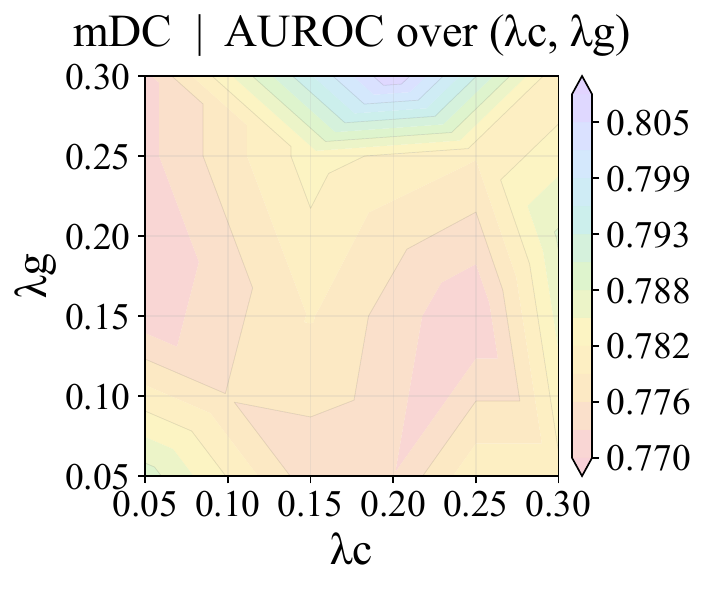}
\includegraphics[width=0.245\textwidth]{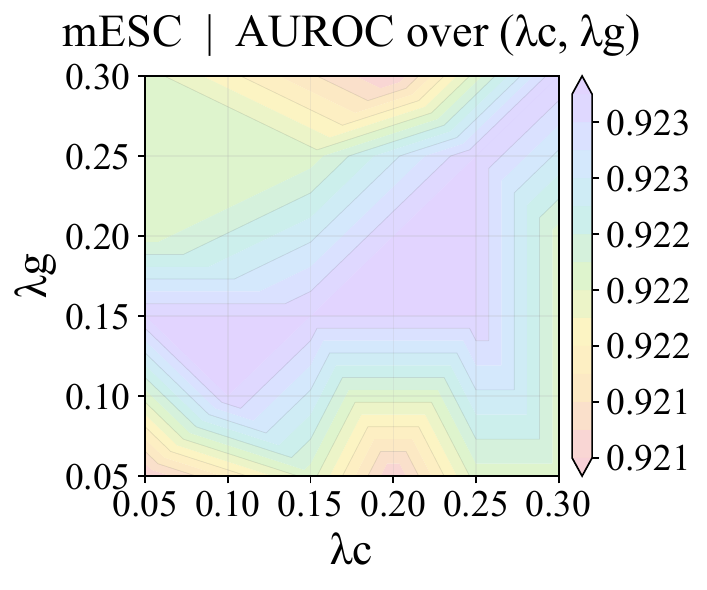}
\includegraphics[width=0.245\textwidth]{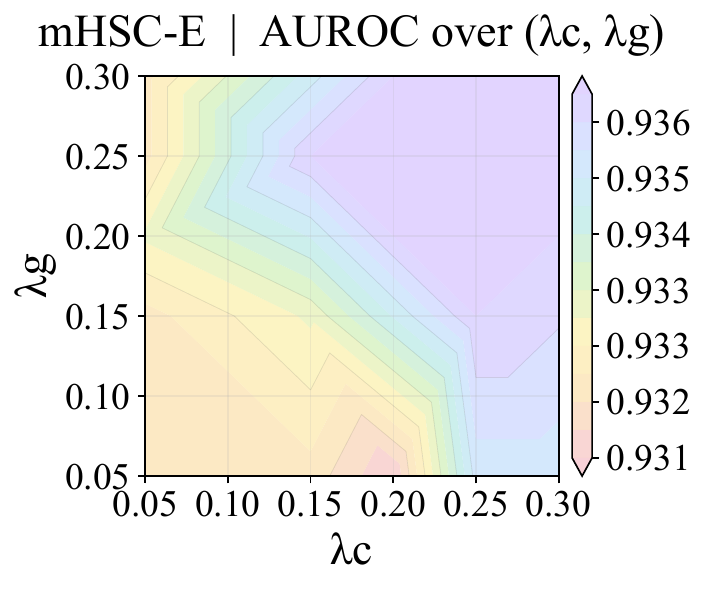}
\includegraphics[width=0.245\textwidth]{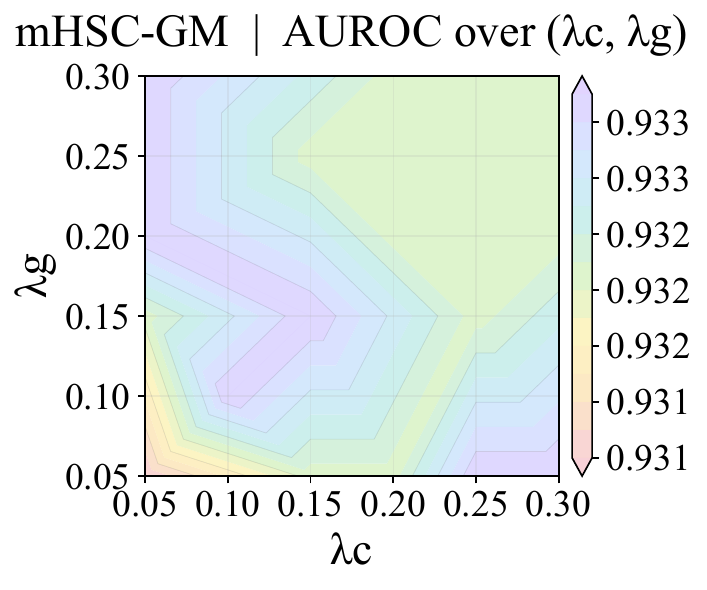}
\includegraphics[width=0.245\textwidth]{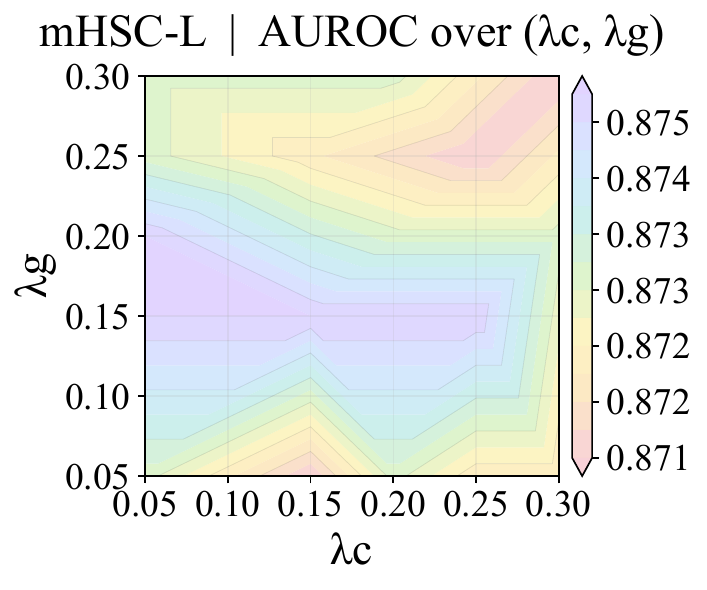}
\caption{AUPRC and AUROC results of sensitivity testing of BRIDGE using a specific network dataset.}
\label{fig:lamda_AUROC}
\end{figure*}

Figure~\ref{fig:lamda_AUROC} shows AUPRC and AUROC over $(\lambda_c,\lambda_g)$ for each cell type. Both metrics vary mildly within a moderate range of $\lambda_c$ and $\lambda_g$, indicating stable behavior under contrastive-weight tuning. Across cell types, the surfaces are generally smooth and do not exhibit sharp peaks, suggesting that performance is not overly sensitive to small perturbations of the contrastive weights. Moderate increases in $\lambda_c$ or $\lambda_g$ can yield incremental improvements, mainly by refining representation consistency. In contrast, extreme values do not provide sustained gains and may degrade performance by disrupting the balance with the link prediction loss, leading to a drop in this regime.

\begin{figure*}[h]
\centering
\includegraphics[width=0.8\textwidth]{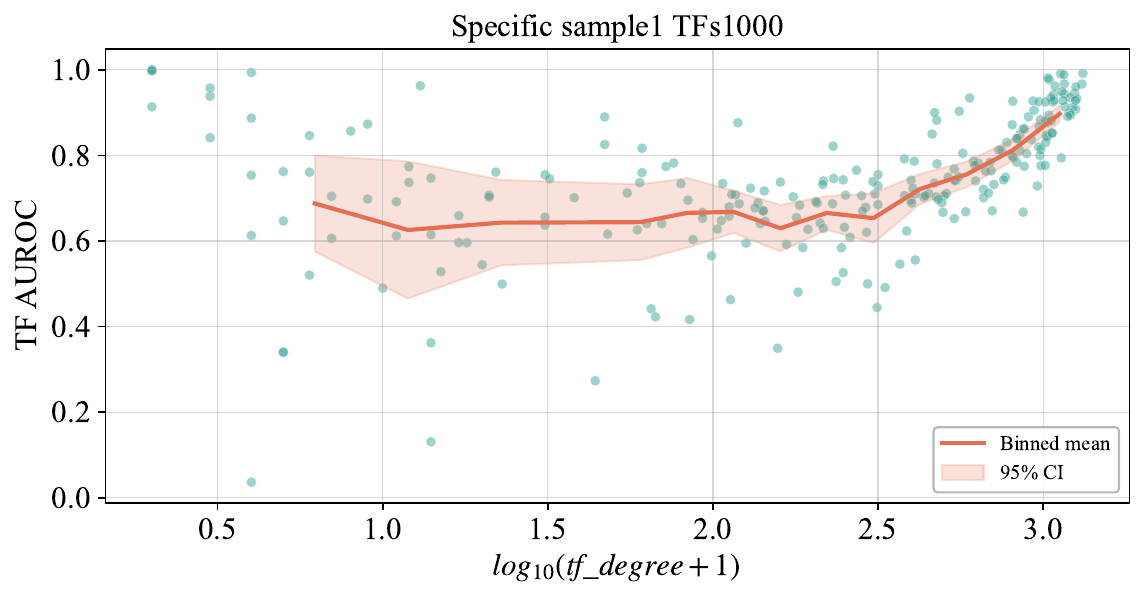}
\caption{Relationship between transcription factor out-degree and TF-level AUROC improvement on the Specific dataset across seven cell types.}
\label{fig:degree}
\end{figure*}

\subsection{Performance Analysis Stratified by TF Degree}
\label{degree}
To evaluate the predictive capability of the proposed model under different degrees of regulatory sparsity, we further conduct a stratified performance analysis based on transcription factor (TF) regulatory scale, aiming to systematically examine the model’s discriminative ability on both long-tail regulatory relationships and highly frequent regulatory interactions. 
Specifically, TFs are stratified according to their out-degree in the regulatory network: TFs regulating no more than 100 target genes are defined as \emph{low-degree} TFs, those regulating between 50 and 200 target genes are defined as \emph{mid-degree} TFs, and TFs regulating more than 200 target genes are defined as \emph{high-degree} TFs. 
Experimentally, we perform evaluation on the \textit{Specific} dataset, where positive and negative samples are relatively balanced. For each of the seven cell types, we load the corresponding trained model and aggregate the results into a unified table, reporting the prediction performance within each degree interval using AUROC and AUPRC.

\begin{table}[h]
\centering
\fontfamily{ptm}\selectfont
\caption{Performance comparison across TF degree groups on the Specific dataset. 
Relative AUPR is defined as $\mathrm{AUPR}/\mathrm{Positive\ Rate}$.}
\begin{adjustbox}{max width=\textwidth}
\begin{tabular}{lccccccccccc}
\hline
Bucket & \#Edges & \#Positive & Positive Rate & \#TFs & Min Degree & Median Degree & Max Degree & AUROC & AUPR & Relative AUPR \\

\hline
High & 45713 & 23526 & 0.514646 & 137 & 204 & 695.0 & 1318 & 0.880543 & 0.892417 & 1.734 \\
Mid  & 14838 & 1225  & 0.082558 & 46  & 54  & 111.0 & 193  & 0.672257 & 0.174898 & 2.119 \\
Low  & 18197 & 185   & 0.010167 & 55  & 1   & 9.0   & 47   & 0.639367 & 0.014774 & 1.453 \\
\hline
\end{tabular}
\end{adjustbox}
\label{tab:tf_degree_bucket}
\end{table}

In terms of predictive performance, the model achieves the best results on high-degree TFs, indicating that under conditions where regulatory interactions are dense and class distributions are relatively balanced, the model can effectively leverage abundant supervisory signals to reliably identify true regulatory relationships. 
For mid-degree TFs, the performance decreases compared to the high-degree group but remains substantially above the random baseline, suggesting that the model retains meaningful discriminative capability under moderate sparsity. 
In contrast, for low-degree TFs, the performance further declines due to the inherent limitation of extremely scarce positive samples, reflecting the objective difficulty of long-tail regulatory prediction.

To further reveal the continuous variation within degree intervals, we additionally analyze the relationship between TF out-degree and TF-level AUROC gain, thereby characterizing the fine-grained performance trend as regulatory scale increases. 
Since AUPRC is strongly influenced by the positive sample ratio, AUROC is adopted in this analysis to more faithfully capture the overall discriminative trend independent of class imbalance(Figure \ref{fig:degree}). 

Overall, the combined results demonstrate that the proposed model consistently attains more stable and pronounced discriminative advantages for TFs with larger regulatory scopes, while still maintaining effective separation capability for mid- and low-degree TFs. 
These findings provide empirical evidence supporting the applicability of the model across regulatory networks with varying degrees of sparsity.

\begin{table}[h]
\centering
\fontfamily{ptm}\selectfont
\caption{AUPRC (mean $\pm$ std) comparison between BER and random edge deletion strategies across different cell types.}
\resizebox{\linewidth}{!}{%
\begin{tabular}{l c c c c c c c}
\hline
 & \multicolumn{7}{c}{\textbf{Cell types}} \\
\cline{2-8}
\textbf{Strategy} 
& \textbf{hESC} 
& \textbf{hHEP} 
& \textbf{mDC} 
& \textbf{mESC} 
& \textbf{mHSC-E} 
& \textbf{mHSC-GM} 
& \textbf{mHSC-L} \\
\hline
BER 
& $0.61 \pm 0.004$ 
& $0.83 \pm 0.002$ 
& $0.14 \pm 0.004$ 
& $0.86 \pm 0.003$ 
& $0.95 \pm 0.002$ 
& $0.95 \pm 0.001$ 
& $0.87 \pm 0.002$ \\

Random edge deletion 
& $0.58 \pm 0.008$ 
& $0.79 \pm 0.007$ 
& $0.12 \pm 0.006$ 
& $0.83 \pm 0.008$ 
& $0.92 \pm 0.004$ 
& $0.93 \pm 0.005$ 
& $0.84 \pm 0.007$ \\
\hline
\end{tabular}%
}

\label{tab:auprc_ber_random}
\end{table}

\begin{figure}[t]
    \centering
    \includegraphics[width=0.99\linewidth]{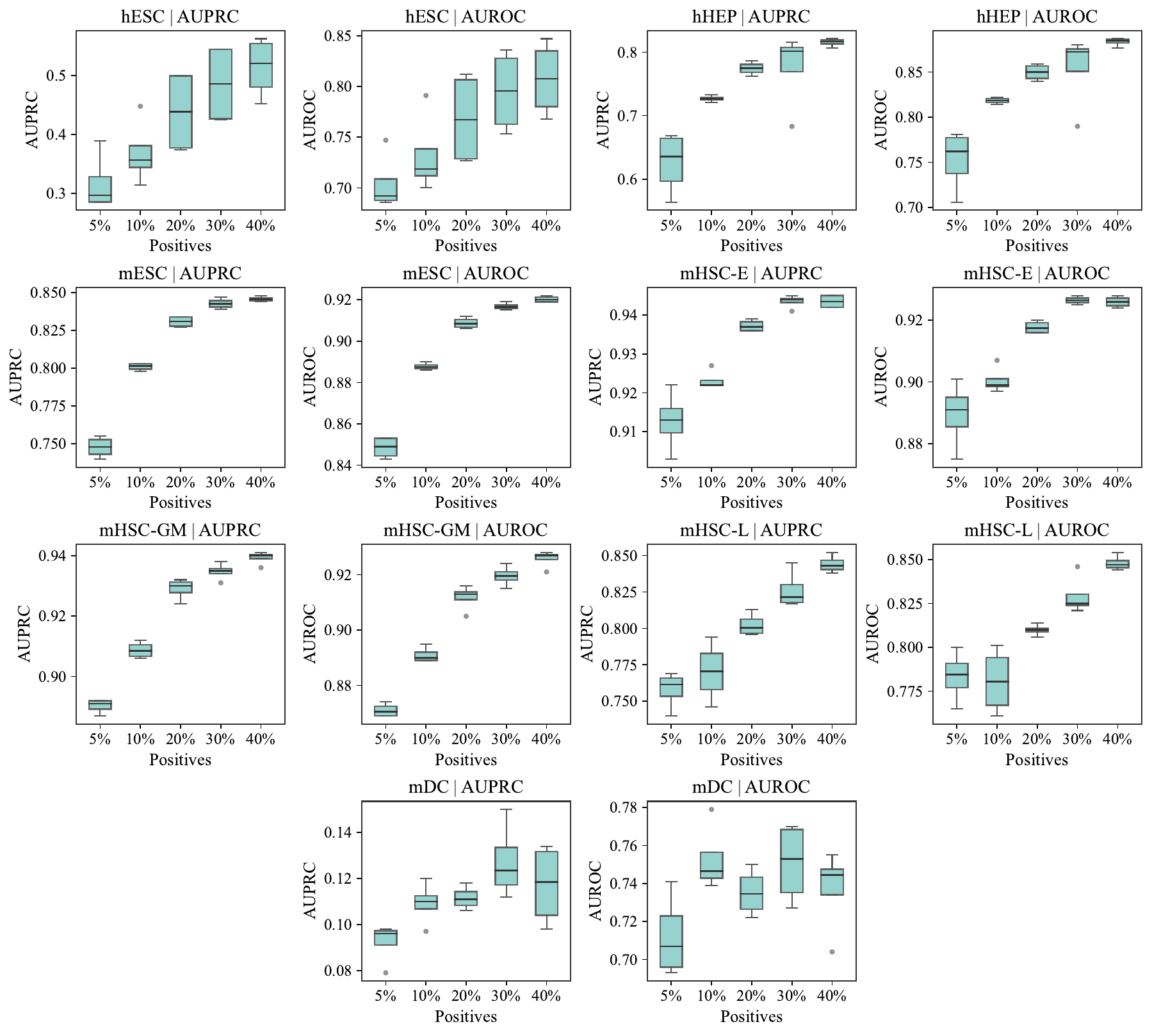}
    \caption{Robustness under Varying Positive Sample Ratios.}
    \label{fig:posrate_robustness}
\end{figure}

\subsection{The Effectiveness and Robustness of the Biological Evidence Refinement for View Augmentation.}
\label{compare}
Using view-enhanced contrastive learning to learn gene representations and applying them to downstream GRN relationship prediction are common and effective approaches. We compare biological evidence refinement for view augmentation (described in Section ~\ref{subsec:BER}) with a random edge removal strategy. BRIDGE is evaluated on seven cell-type datasets under the Specific network setting, where only the view augmentation strategy is replaced and all other components remain unchanged. The results in Table \ref{tab:auprc_ber_random} show that AUPRC is consistently higher with biological evidence refinement than with random edge removal, and the standard deviation is also lower. These findings indicate that biological evidence refinement exploits structural information more effectively and enables more robust representation learning.
The consistent gains and reduced variance support using evidence guided refinement as a principled augmentation for noisy regulatory priors.

\subsection{Positive Sample Ratio Robustness Analysis}
\label{compare2}
In Section \ref{subsec:few-shot}, we have demonstrated that under extremely scarce positive-sample conditions, cross–cell-type transfer learning can effectively alleviate the lack of supervision signals. However, this strategy relies on the availability of a source cell type with sufficient data volume and relatively complete annotations, which may not always be feasible in practical applications. To further evaluate the model’s capability to learn latent regulatory patterns from scratch without external transfer information and under limited positive supervision, we conduct a positive-sample-ratio sensitivity experiment.

Specifically, on the Specific dataset, we construct regulatory networks using the top 1,000 highly variable genes for each of the seven cell types, and progressively reduce the proportion of positive samples to 5\%, 10\%, 20\%, 30\%, and 40\% of the original data. For each positive-sample-ratio setting, the sampling process is independently repeated five times, and the model performance on the test set is evaluated in terms of AUROC and AUPRC. The final results are summarized and visualized using box plots, as shown in Figure \ref{fig:posrate_robustness}.

The results indicate that as the number of positive samples decreases, BRIDGE consistently maintains a high level of predictive accuracy. Notably, even when the positive samples are reduced to only 5\% of the original data, the model is still able to effectively exploit the available supervision and better cope with the challenges posed by severe data imbalance. In addition, our method exhibits relatively small variance across the five random samplings, suggesting a desirable level of stability under different sampling configurations.

\subsection{More Noise Sensitivity Analysis Experiments}
\label{more_result}
\begin{figure*}[t]
\centering
\includegraphics[width=0.32\textwidth]{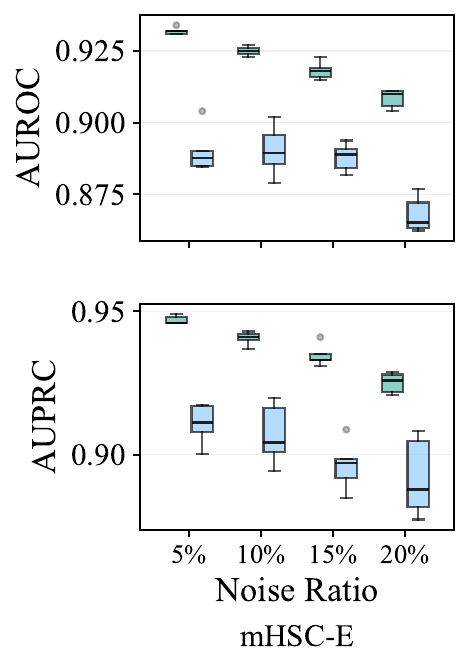}
\includegraphics[width=0.32\textwidth]{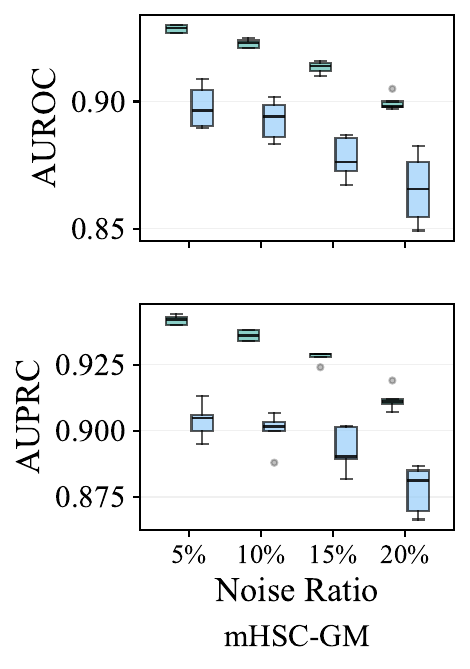}\hfill
\includegraphics[width=0.32\textwidth]{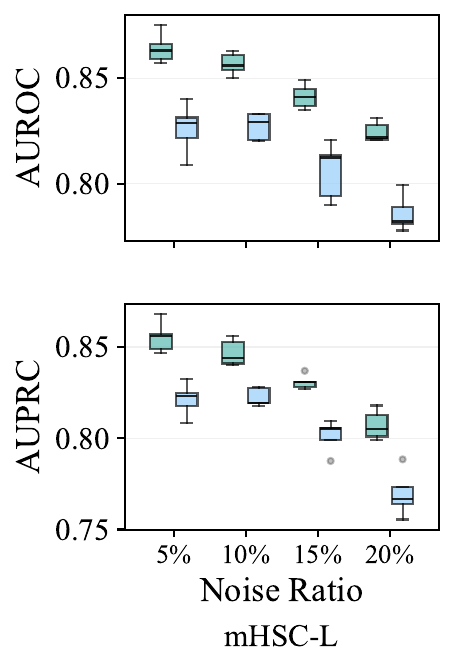}\hfill
\caption{GRN inference metrics (including AUROC above and AUPRC below) at different noise ratios for four cell types in a specific network. The green error bars represent BRIDGE, and the blue ones represent GCLink.}
\label{fig:noisy11}
\end{figure*}

Figure \ref{fig:noisy11} shows the results for three additional cell types (mHSC-E, mHSC-GM and mHSC-L) in the noise sensitivity experiment of Section \ref{noisy}, consistent with previous observations, our model exhibits a stable performance trend across the other three cell types as noise levels increase, demonstrating strong robustness to expression perturbations and label noise across all datasets and outperforming state-of-the-art methods across the board.

Moreover, it is worth noting that the performance degradation under increasing noise remains gradual and well-controlled, without abrupt drops even at higher perturbation levels. This observation suggests that BRIDGE does not rely excessively on a small subset of high-confidence edges, but instead learns distributed and noise-tolerant regulatory representations that are supported by multiple complementary sources of evidence. In particular, the evidence refinement strategy suppresses unreliable or weakly supported interactions, while the gated cross-view integration adaptively regulates the contribution of heterogeneous views, preventing noisy signals from dominating the learning process.
In contrast to methods that are highly sensitive to edge deletion or spurious link injection, BRIDGE effectively mitigates noise propagation through its evidence refinement and gated cross-view integration mechanisms, leading to a more stable embedding space under perturbations. As a result, the inferred regulatory patterns remain consistent across different noise regimes, which is particularly important for real-world single-cell datasets where regulatory networks are inevitably incomplete and contaminated by experimental noise. Overall, these results further confirm that BRIDGE provides a reliable and robust framework for GRN inference under realistic, noisy biological conditions.

\clearpage

\end{appendices}

\end{document}